\documentclass[Journal]{IEEEtran} 

\usepackage{amsthm}
\usepackage{cite}
\usepackage{amsmath}
\usepackage{amssymb}
\usepackage{amsfonts}
\usepackage{amsthm}
\usepackage{graphicx}
\usepackage{fancyhdr}
\usepackage{svg}
\usepackage{bbm}
\usepackage{textcomp}
\usepackage{xcolor}
\usepackage{algorithmicx,algpseudocode}
\usepackage[linesnumbered,ruled]{algorithm2e}
\usepackage{booktabs}
\usepackage{bbold}
\usepackage{kotex}
\usepackage{mwe}
\usepackage{multicol}
\usepackage{multirow}
\usepackage[normalem]{ulem}
\usepackage[bottom]{footmisc}
\usepackage{mwe}
\usepackage{color}

\usepackage{booktabs,multirow}
\usepackage{tabularx}
\usepackage{tikz}
\usepackage{kotex}
\usetikzlibrary{patterns,arrows}
\usepackage{physics}
\usepackage{verbatim}
\usepackage{subcaption}
\usepackage{pifont}
\usepackage[bottom]{footmisc}
\usepackage{circledsteps}
\usepackage{microtype}
\usepackage[export]{adjustbox}
\usepackage{enumitem}

\SetKwRepeat{Do}{do}{while}
\SetKwInOut{Input}{Input}
\SetKwInOut{Output}{Output}

\theoremstyle{definition}

\definecolor{color1}{RGB}{228,26,28}
\definecolor{color2}{RGB}{55,126,184}
\definecolor{color3}{RGB}{77,175,74}
\definecolor{color4}{RGB}{152,78,163}
\definecolor{color5}{RGB}{255,127,0}
\definecolor{color6}{RGB}{241,241,241}
\definecolor{color7}{RGB}{156,156,156}
\definecolor{color8}{RGB}{96,96,96}
\definecolor{color0}{RGB}{162, 20, 47}

\newcommand{\tred}[1]{{\color{black}{#1}}}
\newcommand{\tblue}[1]{{\color{black}{#1}}}

\newcommand*\circled[1]{\Circled[inner color=white, fill color= color0, outer color=color0]{\footnotesize{#1}}} 
\newcommand*\tinycircled[1]{\Circled[inner color=white, fill color= color0, outer color=color0]{\tiny{#1}}}

\newcommand{\cmark}{\ding{51}}
\newcommand{\xmark}{\ding{55}}

\newcommand{\BfPara}[1]{{\noindent\bf#1.}\xspace}

\begin{document}

\title{\fontsize{21}{24}\selectfont A Scalable and Generalizable Pathloss Map Prediction 
}

\author{ 
Ju-Hyung Lee,~\IEEEmembership{Member,~IEEE,} and Andreas F. Molisch,~\IEEEmembership{Fellow,~IEEE,}

\thanks{The authors are with the Ming Hsieh Department of Electrical and Computer Engineering, University of Southern California, 
Los Angeles, USA; email
 \{juhyung.lee, molisch\}@usc.edu. The work was financially supported by NSF projects 2133655 and 2008443. Part of this work was presented at ICASSP 2023 \cite{lee2023large} and Globecom 2023 \cite{lee2023robust}. }
}
\maketitle

\begin{abstract}
Large-scale channel prediction, \textit{i.e.}, estimation of the pathloss from geographical/morphological/building maps, is an essential component of wireless network planning. Ray tracing (RT)-based methods have been widely used for many years, but they require significant computational effort that may become prohibitive with the increased network densification and/or use of higher frequencies in B5G/6G systems. 
In this paper, we propose a data-driven, model-free pathloss map prediction (PMP) method, called PMNet. PMNet uses a supervised learning approach: it is trained on a limited amount of RT (or channel measurement) data and map data. Once trained, PMNet can predict pathloss over location with high accuracy (an RMSE level of $10^{-2}$) in a few milliseconds.
We further extend PMNet by employing transfer learning (TL). TL allows PMNet to learn a new network scenario quickly ($\times 5.6$ faster training) and efficiently (using $\times 4.5$ less data) by transferring knowledge from a pre-trained model, while retaining accuracy. 
Our results demonstrate that PMNet is a scalable and generalizable ML-based PMP method, showing its potential to be used in several network optimization applications.
\end{abstract}

\begin{IEEEkeywords}
Pathloss map prediction, ray tracing, channel measurement, machine learning, computer vision, transfer learning, network optimization, digital twin, 6G.
\end{IEEEkeywords}

\section{Introduction} \label{sec:introduction}
\tblue{Digital twin (DT) technology is emerging as a key enabler for the artificial intelligence (AI) and machine learning (ML)-driven design, simulation, and optimization of 6G systems \cite{lin2023digital, alkhateeb2023realtime}. A DT is a dynamic, digital replica of a real-world network environment, providing real-time, accurate reflections of physical network scenarios. 
However, implementing DT is challenging in 6G networks, which are characterized by increased deployment density, complex distributed architectures, and high-frequency operation in millimeter wave (mmWave) and terahertz (THz) bands.}

Individually and taken together, these developments necessitate dramatically faster large-scale channel prediction methods.\footnote{The word ''channel prediction" is often used for two different problems: (i) computation of the propagation channel at a particular location based on maps of the environment, and (ii) temporal prediction of the channel (often for a mobile device moving on a trajectory), based on measurements in the immediate past. This paper only considers the former case.} 
Since traditional ray tracing (RT) tools are too slow for the repeated runs required in such DT implementation processes, there is a strong need for new, accurate, and fast methods for channel prediction over a large-scale area (\textit{e.g.}, campus or city-map scale).

Several works have addressed this need by channel prediction using powerful ML techniques. These works use ground-truth channel data (from RT simulations or real channel measurements/soundings campaigns) to train neural networks (NNs). This eventually provides an accurate and fast prediction of channel information (\textit{e.g.}, received power, delay, angles, and so on) for a certain area, a technique called ML-based site-specific radio propagation modeling.

Still, these ML-based approaches use supervised learning, meaning they are trained to solve a specific network scenario with a certain labeled dataset. In other words, the models may need to be rebuilt for a new network scenario, \textit{e.g.}, different map scales, environmental aspects, and/or network configuration - a process that can be time-consuming and expensive.
This creates a need for a method that can furthermore transfer knowledge of propagation channels across different network scenarios and environments.

\begin{figure}[t!]
    \centering
    \includegraphics[width=1.\linewidth]{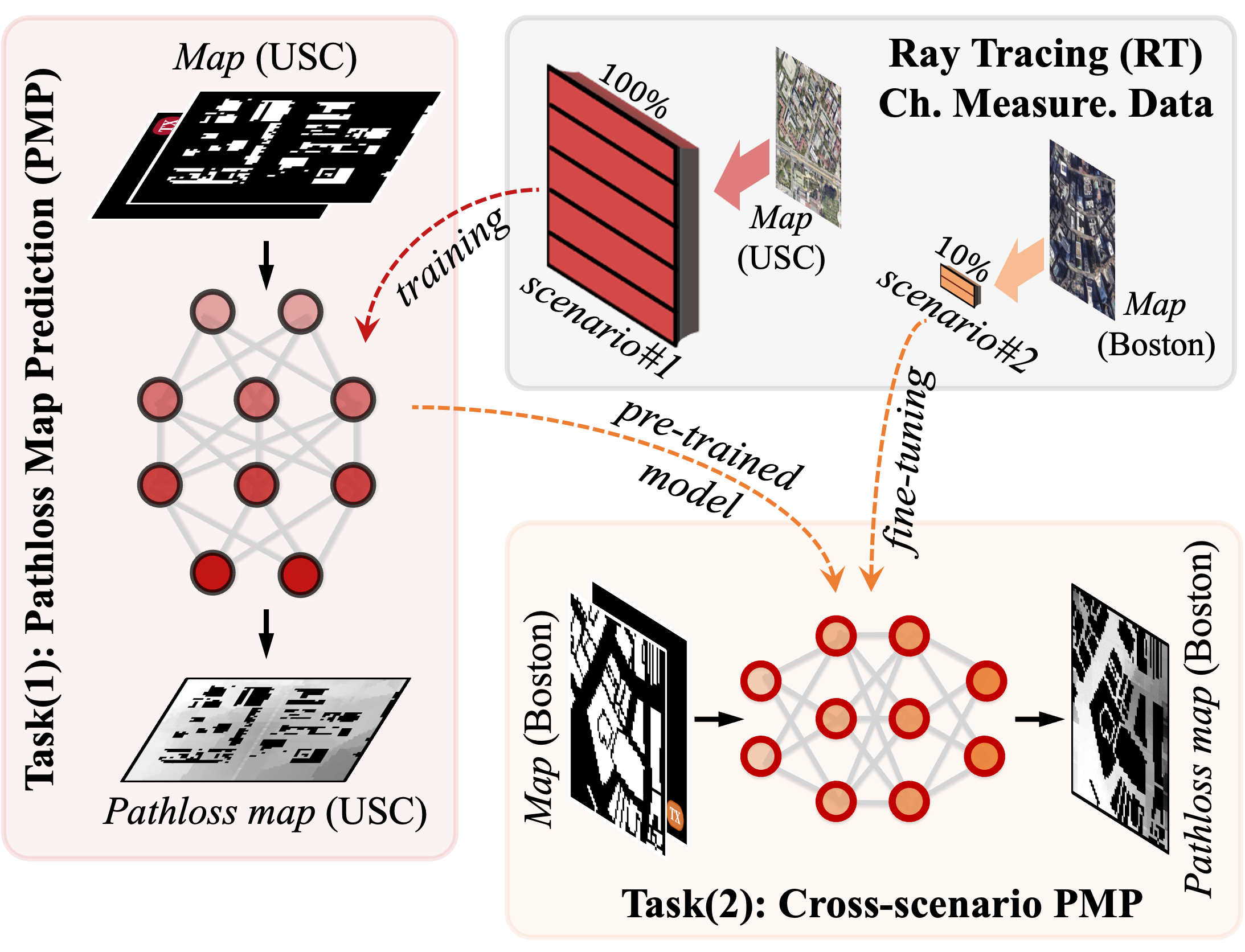}
    \caption{\tblue{Overview of the pathloss map prediction (PMP) task and the cross-scenario PMP. The input Map feature includes the transmitter (TX) location.}}
    \label{fig:overview_PMP}
    \vspace{-1.0em}
\end{figure}

\subsection{Related Works}
Due to the high cost and complexity of field measurements with channel sounders, most cellular deployment planning has long replaced channel measurements with \emph{electromagnetic (EM) simulation-based} approaches, such as RT \cite{valenzuela1993ray,kim1999radio} and ray launching \cite{WirelessInsite} simulation.\footnote{Throughput this paper, we will use the terms "measurements", "ray tracing", and "ray launching" interchangeably to refer to a suitable method for finding a site-specific ground-truth for the pathloss.} Over the past 30 years, the efficiency and accuracy of RT have improved significantly \cite{degli'Esposti2007}, thanks to the prevalence of GPUs (graphic processing units) that efficiently facilitate RT tasks.

However, due to the factors mentioned above (such as the need for more detailed environmental consideration at higher frequencies and the need for fast simulations with higher deployment density), RT simulations are too computationally intensive for large-scale network deployment in 6G systems. As a result, simplified \emph{model-based} approaches like the dominant path model \cite{dominant}, or fine-tuning of generic pathloss models (\textit{e.g.}, 3GPP path gain model) with limited measurement data \cite{voronoi, survey_radiomap} have been proposed over the years. However, these approaches have found only limited acceptance by network operators due to their insufficient accuracy in predicting the propagation characteristics of signals in complex environments.

\tblue{
In recent years, supervised ML has been applied to solve a variety of challenging problems in wireless communication, including channel measurement/prediction for 6G networks. Such an \emph{ML-based} approach can be trained on a map of the environment (topology/morphology) and a relatively small set of measurement data to learn how to provide a virtual replica (\textit{e.g.}, DT) of a large-scale network environment in real-time while accurately modeling the behavior of channel characteristics.

On the one hand, models like WiNeRT \cite{WiNeRT} and NeRF2 \cite{NeRF2} are specifically developed to predict detailed channel information (\textit{e.g.}, power, delay, and angle information) of each multi-path component (MPC) between TX and receiver (RX) with the input of detailed information, including spatial configuration and wireless configuration parameters. 
These models are particularly well-suited for applications in small-scale indoor areas, where high-detailed channel prediction is required (\textit{e.g.}, indoor sensing).

On the other hand, models like RadioUNet \cite{RadioUNet} and FadeNet \cite{Fadenet} aim to predict the path gain, received power, or coverage for TX-RX in a given area with the input of a building map. 
These models are designed for large-scale channel prediction, where fast operation is essential (\textit{e.g.}, network optimization). 

In particular, several state-of-the-art works, such as Agile \cite{enes2023agile}, PPNet \cite{kehai2023deep}, and PMNet \cite{lee2023large}, are pushing the boundaries of predictive accuracy and computational efficiency for large-scale channel prediction, as evidenced by their performance in ML competitions such as the \emph{RadioMap Challenge} (see details in \cite{RadiomapChallenge}). This highlights the applicability and importance of large-scale channel prediction in evolving wireless network optimization, which aligns with our research direction.
}



\subsection{Contributions}
This paper proposes a scalable and generalizable channel prediction approach specifically designed for large-scale channel prediction, called PMP task. Our contributions can be summarized as follows:
\begin{itemize}
\item We design a PMP-oriented NN architecture, called PMNet, by leveraging computer-vision techniques, generating highly accurate channel prediction results for a given map in few milliseconds. PMNet achieves the best channel prediction accuracy compared to two baselines: a model-based scheme (\emph{3GPP-UMi} model \cite{3gpp2018pathloss}) and another ML-based scheme (\emph{RadioUNet} \cite{RadioUNet}) (see \textbf{Table \ref{table:comparison_baseline}} in Sec. \ref{sec:pmnet}) and also in different PMP datasets. PMNet achieved \emph{1st}-rank in the ICASSP 2023 Radio Map Challenge \cite{RadiomapChallenge}.\footnote{In this competition, PMNet demonstrated its high accuracy in the PMP task even on a different dataset \cite{RadioMapSeer} with a different map scale and network configuration, and generated by a different RT simulation tool, \textit{i.e.}, \emph{WinProp}.}

\item We build three sets of real-world channel measurement datasets using a RT simulation tool, \textit{i.e.}, \emph{Wireless Insite}, for training and evaluation, which reflects different network scenarios (\textit{e.g.}, different map scale, environment, and network configuration) in two different light urban environments (the USC and UCLA campuses) and a metropolitan area (the Boston area), see \textbf{Table \ref{table:dataset}} in Sec. \ref{sec:dataset}.

\item We propose a method of predicting pathloss in unseen network scenarios by using transfer learning (TL) with a pre-trained model. 
We prepare three pre-trained models for TL: VGG16 \cite{vgg16} and two pre-trained PMNet models trained with 3GPP prediction results and RT simulation results, respectively, and quantitatively and qualitatively evaluate their accuracy (see \textbf{Table \ref{table:comparison_TL}} and \textbf{Fig. \ref{fig:comparison_TL}} in Sec. \ref{sec:transferlearning}). 

\item We empirically demonstrate that our PMNet pre-trained model has generalization capability for different network scenarios, adjusting to new network scenarios $\times 5.6$ faster and using $\times 4.5$ less data than a baseline model without TL, while still achieving high accuracy of an RMSE of $10^{-2}$ level (see \textbf{Fig. \ref{fig:comparisn_TL_val}} and \textbf{Table. \ref{table:requiredStep}} in Sec. \ref{sec:transferlearning}).

\item \tred{We release source code for the experiments to promote reproducible ML research in wireless communication.\footnote{{https://github.com/abman23/PMNet}}}
\end{itemize}

\subsection{Paper Organization}
The rest of the paper is organized as follows: 
Sec.~\ref{sec:background} presents the background on two important concepts: (1) \emph{ray tracing simulation}, which is used to generate ground-truth channel information for training and evaluation; and (2) \emph{transfer learning}, which enables us to transfer the knowledge learned from a source task/dataset   to a new task/dataset (\textit{e.g.}, unseen network scenario). 
After introducing our dataset based on real geographical maps in Sec.~\ref{sec:dataset}, Sec.~\ref{sec:pmnet} introduces the PMP task and our proposed NN architecture (PMNet) for this channel prediction task. We also present the training and evaluation process, as well as simulation results.  
Then, Sec.~\ref{sec:transferlearning} presents our approach for efficiently learning and predicting channels in unseen network environments by transferring the pre-trained knowledge from other networks. We provide extensive experimental results and quantitative and qualitative performance analysis, followed by concluding remarks in Sec.~\ref{sec:conclusion}.

\textit{Notation:}
Throughout this paper, we use the normal-face font to denote scalars and the boldface font to denote vectors.
We use $P(\cdot)$ and $P(\cdot|\cdot)$ to represent a marginal probability distribution and conditional distribution, respectively.
We also use $\|\cdot\|$ to denote the $L^2$-norm, which is an Euclidean norm.
$\mathcal{N}(\mu,\sigma)$ denotes the normal distribution with mean $\mu$ and  standard deviation $\sigma$.



\section{Background} \label{sec:background}
\subsection{Pathloss} \label{sec:pathloss}
The link gain between a TX at location $\vb*{q}_{\mathrm{TX}}$ and an RX at location $\vb*{q}_{\mathrm{RX}}$ at time $t$ and frequency $f$ can be expressed as follows:
\begin{equation}
    |h(t,f,\vb*{q}_{\mathrm{TX}},\vb*{q}_{\mathrm{RX}})|^2 = \frac{P_{\mathrm{RX}}(t,f,\vb*{q}_{\mathrm{RX}})}{P_{\mathrm{TX}}(t,f,\vb*{q}_{\mathrm{TX}})}
\end{equation}
where $P_{\mathrm{RX}}$ and $P_{\mathrm{TX}}$ are received and transmitted power, respectively. This link gain includes the effects of antenna gains at TX and RX; when isotropic antennas are used, it becomes identical to the channel gain. 
It exhibits variations in time and/or location due to small-scale fading, shadowing, and large-scale distance changes. Averaging over small-scale fading removes (under certain circumstances, see \cite[Ch. 7]{molisch2023wireless}) the dependence on frequency and time, providing the {\em path gain} (PG) that can be written as a function of only the large-scale distance changes:
\begin{equation}
    \mathrm{PG}(\vb*{q}_{\mathrm{TX}},\vb*{q}_{\mathrm{RX}}) = \dfrac{1}{T_{\mathrm{S}}} \dfrac{1}{B_{\mathrm{S}}} \int\limits_{T_{\mathrm{S}}}\int\limits_{B_{\mathrm{S}}} |h(t,f,\vb*{q}_{\mathrm{TX}},\vb*{q}_{\mathrm{RX}})|^2 \,df\,dt.
\end{equation}
Here, $T_{\mathrm{S}}$ and $B_{\mathrm{S}}$ denote the stationary-time and -bandwidth, respectively.
The path gain can be represented as the sum of the powers of the $N$ MPCs, as discussed further in Sec. \ref{sec:channel messurement}. For later reference, we note that the pathloss is the inverse of the path gain (or the sign-flipped value when expressed in dB).  

\subsection{Ray Tracing (RT) Simulation}
RT is an approximate method for modeling the propagation of electromagnetic waves in wireless communication scenarios. It works by tracing the paths of individual rays as they propagate through the environment, whose features are represented in a geographical database. The rays are reflected, deflected, and scattered by the objects in the environment, with the various interaction processes computed according to high-frequency approximations, namely (most commonly) Snell's laws for specular reflection and transmission, uniform theory of diffraction (UTD) for diffraction, and Kirchhoff scattering theory for diffuse scattering \cite[Ch. 4]{molisch2023wireless}.\footnote{RT can be implemented via image-theory-based RT, or as ray launching. We will henceforth use the expression RT for both those methods.} In this paper, we employ a commercial RT tool, \emph{Wireless Insite} from Remcom \cite{WirelessInsite} for all RT simulations, both because of its user-friendliness and the fact that its accuracy has been compared against a number of channel sounder measurements \cite{kim1999radio, fuschini2008analysis, rautiainen2002verifying}. 

RT can be used to predict channel information, such as received signal strength, delay, and angles, in a variety of wireless environments, both indoor and outdoor.  The accuracy of RT simulations depends on various factors, such as the complexity of the environment, the accuracy of the geographical database, and the carrier frequency. The channel information obtained from the RT can be utilized, {\em inter alia}, for various network optimization tasks, including base station (BS) deployment planning, BS parameter optimization, as well as beam management and localization.

\subsection{Transfer Learning (TL)} \label{sec:transferlearning}
TL is a ML technique that involves reusing a pre-trained model on a new task. This is particularly useful when there is limited data available for the new task or when the new task is similar to a task that has already been learned.
For example, a model pre-trained on image classification can be used for object detection or semantic segmentation.

One of the most popular pre-trained models is VGG16 \cite{vgg16}, which is trained on more than a million images from the ImageNet database for image classification. VGG16 has been reused to improve the performance of a wide variety of tasks, such as semantic segmentation \cite{long2015fully} and object detection \cite{ren2015towards}.

However, it is important to note that the effectiveness of TL depends on the \emph{similarity} between the pre-trained task and the target task. 
The transferability of deep feature representations decreases as the discrepancy between the pre-trained task and the target task increases \cite{yosinski2014transferable}. In other words, the further apart the task is, the less transferable the knowledge. One example is catastrophic forgetting, which is a phenomenon that can occur when fine-tuning a pre-trained model on a new task, resulting in a loss of previously acquired knowledge \cite{kirkpatrick2017}.

Research has shown that well-generalized models, particularly those with excellent pre-training performance \cite{kornblith2019better}, have the potential to require minimal fine-tuning or even none at all (\textit{e.g.}, zero-shot learning) for new tasks \cite{huh2016makes}. These suggest the importance of selecting a pre-trained model suitable for the target task.

\section{Dataset} \label{sec:dataset}
\begin{figure}[h!]
\centering
\begin{subfigure}[t]{.48\linewidth}
\centering
\includegraphics[width=\linewidth]{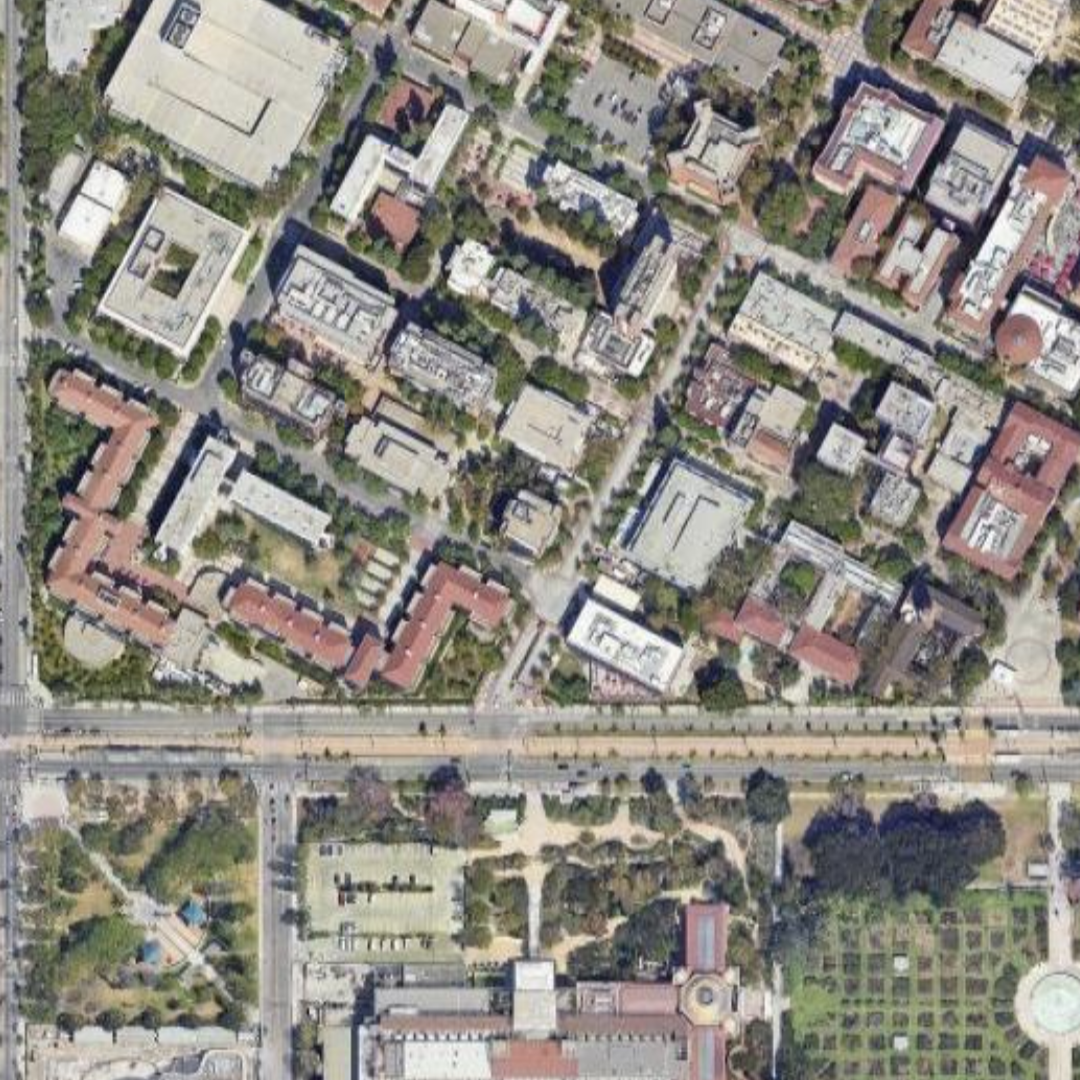}
\caption{USC campus (Map)}
\label{map:USC}
\end{subfigure}
\hspace*{.025\linewidth}%
\begin{subfigure}[t]{.48\linewidth}
\centering
\includegraphics[width=\linewidth]{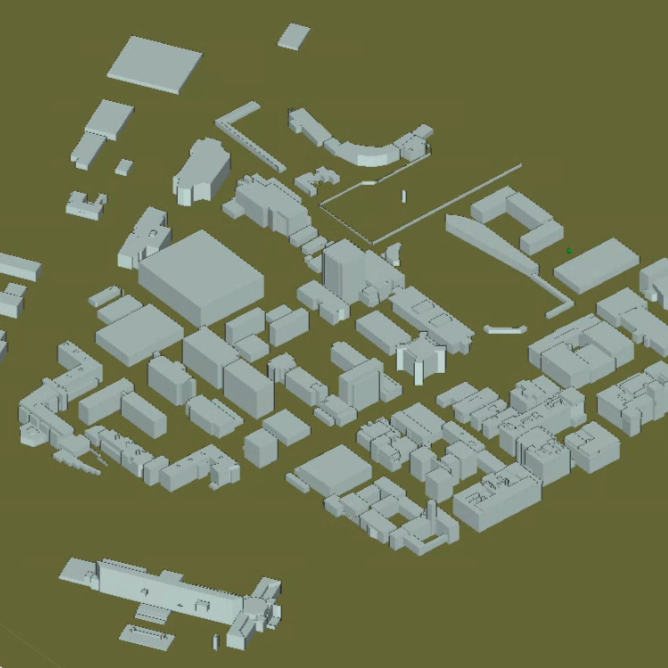}
\caption{USC campus (Geometry in \emph{Wireless Insite})}
\label{geometry:USC}
\end{subfigure}
\\
\vspace*{.025\linewidth}%
\begin{subfigure}[t]{.48\linewidth}
\includegraphics[width=\linewidth]{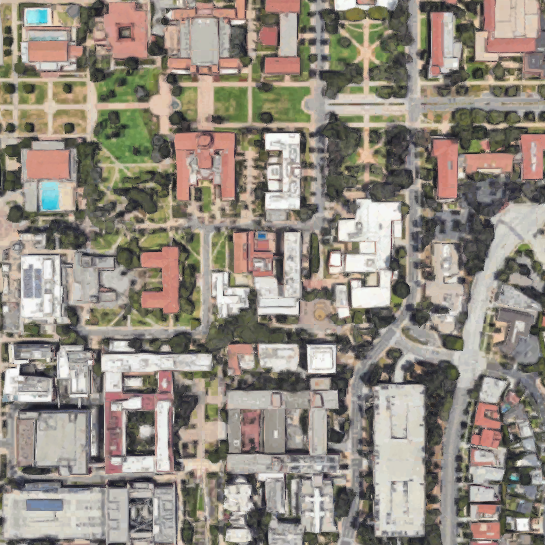}
\caption{UCLA campus (Map)}
\label{map:UCLA}
\end{subfigure}
\hspace*{.025\linewidth}%
\begin{subfigure}[t]{.48\linewidth}
\centering
\includegraphics[width=\linewidth]{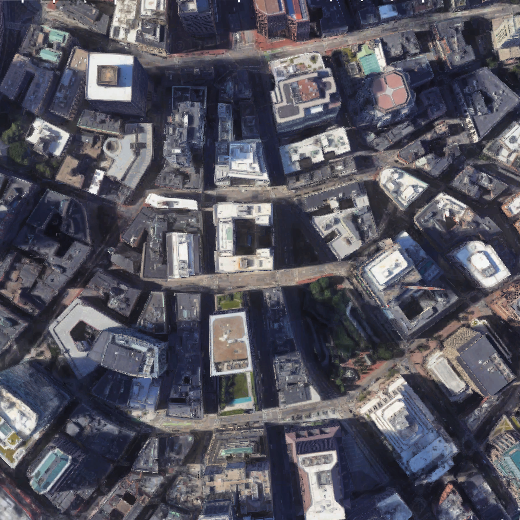}
\caption{Boston (Map)}
\label{map:Boston}
\vspace{-1.0em}
\end{subfigure}
\caption{Map of USC, UCLA, and Boston used in RT simulation. Fig.~\ref{map:USC} is imported and converted to Fig.~\ref{geometry:USC}. The ground-truth pathloss map over the USC campus is then obtained using \emph{Wireless Insite} RT simulation and pre-processing (\textit{e.g.}, interpolation, gray conversion, and data augmentation).}
\label{fig:datatset}
\end{figure}

In this section, we discuss the dataset preparation process for our pathloss map datasets, reflecting real-world network scenarios in USC, UCLA, and Boston areas.

We obtained the ground-truth channel measurement data using the commercial RT tool \emph{Wireless Insite} \cite{WirelessInsite}, which takes into account the geographical and morphological features of the propagation environment. We then pre-processed the data (\textit{e.g.}, interpolation and data augmentation) to prepare the ground-truth pathloss map.

\subsection{Channel Measurement} \label{sec:channel messurement}

\subsubsection{RT simulation}
As discussed in Sec. \ref{sec:pathloss}, RT emulates the behavior of each MPC between TX and RX, following physical principles including the free-space power loss and interaction with different interacting objects (IOs). This allows us to compute for each MPC the information of complex amplitude $a$, directions of departure $\Omega$ and arrival $\Psi$, and delay $\tau$. 
The contribution of $m$-th MPC can be expressed as \cite{steinbauer2001double}:
\begin{align}
    h_{m}(t,\tau, \Omega, \Psi) = a_{m} \delta(\tau - \tau_{m}) \delta(\Omega - \Omega_{m}) \delta(\Psi - \Psi_{m}),
\end{align}
where the dependence of $\Omega,\Psi,\tau,a$ on $t$ is not written explicitly on the r.h.s. 
The sum of contributions from all MPCs is given by 
\begin{align}
    h(t,\tau, \Omega, \Psi) =  \sum_{m=1}^{N} h_{m}(t,\tau, \Omega, \Psi).
    \label{eq:ddimpulseresponse}
\end{align}
Since $\Omega,\Psi,\tau,|a|$ are constant over a stationarity-time and bandwidth, while $arg(a)$ varies over many periods of $2\pi)$, and assuming isotropic antennas at TX and RX (so that $\Omega,\Psi$ do not matter), the path gain  averaged over the small-scale fading can be computed from (2) as
\begin{align}
    \mathrm{PG} =  \sum_{m=1}^{N} | h_{m}(\tau, \Omega, \Psi)|^2 = \sum_{m=1}^{N} |a_{m} |^2. \label{eq:pathgain}
\end{align}
Note that our pathloss map uses the information of path gain (in [dB]) while other information on angles and delay is not needed (though this information can be used for further applications, \textit{\textit{e.g.}}, beamforming algorithms). 

Thus, $P_{\mathrm{RX}}$ (in [dBm]) can be expressed as a function of $P_{\mathrm{TX}}$ (in [dBm]) as follows:
\begin{align}
    P_{\mathrm{RX}} = P_{\mathrm{TX}} + \mathrm{PG}. \label{eq:rxpower}
\end{align}
\tblue{
Note that we set $ P_{\mathrm{TX}}=0$ [dBm] in our RT dataset to simplify the analysis, which makes $P_{\mathrm{RX}}$ in [dBm] equal to $\mathrm{PG}$ in [dB]. 
}

To generate a ground-truth (labeled) dataset that simulates real-world network scenarios, we conduct \emph{Wireless Insite} RT simulations on the geographical and morphological maps of the University of Southern California (USC) campus, the University of California, Los Angeles (UCLA) campus, and the Boston area. Both campus areas are in Los Angeles, CA, and exhibit a (light) urban build-up, with most buildings being five stories or less (with a few high-rises interspersed), gaps between buildings along the street canyons, and some open squares. The Boston area is in downtown of Boston, MA. It is a metropolitan area with multiple high-rises; its streets are {\em not} arranged along a rectangular grid.
Each dataset has different network configurations and environmental characteristics (\textit{e.g.}, map scale, and geographical features, such as vegetation).
See Fig. \ref{fig:datatset} and Table \ref{table:dataset} for more details.\footnote{  
It is worth noting that the simulations are performed at the sub-6 GHz band, which is the most widely used cellular band. Similar simulations can be performed in other frequency bands, such as the mmWave and THz bands, with minor adjustments to the parameters. However, at those high frequency bands, geographical data bases with higher resolution might be required for comparable accuracy.
}

We stress that the goal of our work is the correct prediction of ''ground-truth" pathloss by ML techniques. The pathloss obtained from the RT simulations might deviate from measured values due to inaccuracies of the database or inherent approximations of RTs. However, such deviations are irrelevant to the assessment of our ML methods, since they only impact what is used as ''ground-truth" and not the prediction process itself. In other words, if the ground-truth is more accurate (similar to measurement results), our prediction inherently becomes more accurate as well. 

\begin{table*}[!ht]   
  \centering
  \caption{Parameters of USC, UCLA, and Boston datasets.}
  \resizebox{2.0\columnwidth}{!}{\begin{minipage}[t]{1.\textwidth}
  \centering
  \begin{tabular} {l  c c c}
\toprule[1.5pt]
\multirow{2}{*}{\textbf{\textit{Parameter}}} & \multicolumn{3}{c}{\textbf{\textit{Dataset}}}  \\
\cmidrule(lr){2-4} 
& USC  & UCLA  & Boston \\
\cmidrule(lr){1-1} \cmidrule(lr){2-2} \cmidrule(lr){3-3} \cmidrule(lr){4-4}
Map scale
& $880 \times 880$ {[}m$^2${]} 
& $760 \times 760$ {[}m$^2${]} 
& $553 \times 553$ {[}m$^2${]} \\  
Cropped map scale (per pixel)
& $221 \times 221$ {[}m$^2${]} ($0.86 \times 0.86$ {[}m$^2${]}) 
& $225 \times 225$ {[}m$^2${]} ($0.88 \times 0.88$ {[}m$^2${]}) 
& $187 \times 187$ {[}m$^2${]} ($0.73 \times 0.73$ {[}m$^2${]}) \\  
Terrain & \cmark & \cmark & \cmark \\
Buildings & \cmark & \cmark & \cmark \\
Foliage & \xmark & \xmark & \cmark \\
\cmidrule(lr){1-1} \cmidrule(lr){2-2} \cmidrule(lr){3-3} \cmidrule(lr){4-4}
Carrier frequency & $2.5$ {[}GHz{]} & $3.0$ {[}GHz{]} & $3.0$ {[}GHz{]} \\
Transmit power & $0$ {[}dBm{]} & $0$ {[}dBm{]} & $0$ {[}dBm{]} \\
TX antenna type\footnote{\tblue{Isotropic and half-wave dipole antennas provide almost identical radiation patterns within a certain angular extent. MPC induced outside of the angular extent does not contribute significantly to the link.}} & Isotropic (vertical) & Half-wave dipole (vertical) & Half-wave dipole (vertical) \\ 
\cmidrule(lr){1-1} \cmidrule(lr){2-2} \cmidrule(lr){3-3} \cmidrule(lr){4-4}
Total \# of data/scene & $4754$ & $3776$ & $3143$ \\ 
\bottomrule[1.5pt]
\vspace{-1.0em}
\end{tabular}

  \label{table:dataset}
  \end{minipage}}
  \vspace{-1.0em}
\end{table*}

\subsubsection{3GPP model} \label{sec:3gpp model}
The 3GPP 38.901 channel model \cite{3gpp2018pathloss} (henceforth simply called the ''3GPP model" for conciseness) is a widely used model for wireless system standardization that claims validity for frequencies spanning from $0.5$ to $100$ [GHz]. It utilizes TDL or clustered delay line (CDL) to model the double-directional impulse response in \eqref{eq:ddimpulseresponse}. This model defines clusters as collections of paths that share the same delay but have slightly different angles. Cluster delays can either be deterministic or random, with cluster power decreasing as the delay increases. 

\tblue{For the purposes of this paper, we only consider the 3GPP modeling of the pathloss, 
which follows the classical $\alpha-\beta$ model 
\begin{equation}
    \mathrm{PL}_{\alpha-\beta}(d) = 10 \alpha \log_{10}(d) + \beta + S,
\end{equation}
where $S \sim \mathcal{N}(0, \sigma_{S})$ is a lognormally distributed random variable (with variance $\sigma_{S}$) representing the shadow fading, and $\alpha$, $\beta$, and $\sigma$ are parameters of the model that are based on measurement campaigns and that are different in different environments. Important for our later discussions, those parameters are also different depending on whether an unobstructed optical line of sight (LoS) exists between TX and RX or not.}

Specifically, for urban environments, the following describes the path gain:
\begin{align}
& \mathrm{PG}_{\mathrm{UMi-LoS}} = 
\begin{cases}
    \mathrm{PL}_1, \ (10 [\mathrm{m}] \leq d_{\mathrm{2D}} \leq d_{\mathrm{BP}}) \\
    \mathrm{PL}_2, \  (d_{\mathrm{BP}} \leq d_{\mathrm{2D}} \leq 5 [\mathrm{km}])
\end{cases}
\end{align}
\begin{align}
\ \mathrm{PG}_{\mathrm{UMi-NLoS}} = \max(& \mathrm{PG}_{\mathrm{UMi-LoS}}, \mathrm{PL}_3),  \\ 
& \ \ \ \ \ \ (10 [\mathrm{m}] \leq d_{\mathrm{2D}} \leq 5 [\mathrm{km}])  \nonumber
\end{align}
where the  two-dimensional $xy$-distance is $d_{\mathrm{2D}}$ and the three-dimensional $xyz$-distance is $d_{\mathrm{3D}}$, 
\begin{eqnarray}
 \mathrm{PL}_1 & = & 32.4 + 21 \log_{10}(d_{\mathrm{3D}}) + 20 \log_{10}(f_c), \\ \mathrm{PL}_2 & = & 32.4 + 40 \log_{10}(d_{\mathrm{3D}}) + 20 \log_{10}(f_c) \nonumber \\ & & - 9.5 \log_{10}((d_{\mathrm{BP}})^2 + (h_{\mathrm{BS}}-h_{\mathrm{UT}})^2), \nonumber 
\end{eqnarray}
\begin{eqnarray}
 \mathrm{PL}_3 & = & 22.4 + 35.3 \log_{10}(d_{\mathrm{3D}}) + 21.3 \log_{10}(f_c) - \nonumber \\ & & 0.6(h_{\mathrm{UT}} - 1.5).   \nonumber  
\end{eqnarray}
Here, the breakpoint distance is $d_{\mathrm{BP}}=2 \pi h_{\mathrm{BS}} h_{\mathrm{UT}} f_c / c$ where $f_c$ is the center frequency in [Hz] and $c=3.0 \times 10^8$[m/s] is the speed of light. The antenna heights at the TX (\textit{e.g.}, base station), $h_{\mathrm{BS}}$, and the RX (\textit{e.g.}, user terminal), $h_{\mathrm{UT}}$, are set to $1.5$ [m] and $10$ [m], respectively. Note that the model differs for LoS and non-LoS (NLoS) situations.  

This model is employed as one of our baselines for the prediction (see Sec. \ref{sec:pmnet_simulation}). While the 3GPP model also models shadowing, it incorporates it as {\em stochastic} variations that cannot be related to particular map features; we therefore omit them for the purposes of this paper.

\subsection{Pre-processing} \label{sec:preprocessing}
The raw numeric data from the RT simulation is pre-processed using \emph{gray conversion} and \emph{interpolation} methods to generate the ground-truth pathloss map, \emph{data augmentation} methods to create an increased amount of labeled data, and \emph{sampling} methods to divide them into training and testing sets.

\BfPara{Gray conversion}\quad
\tblue{
To generate the pathloss map, we begin by converting the received power $P_{\mathrm{RX}}$ (in [dBm]) (or the path gain $\mathrm{PG}$ in [dB]) into grayscale between $1$ and $255$ using \emph{Min-Max} normalization, with the minimum value of $-254$ [dBm] and the maximum value of $0$ [dBm]. 
While the upper value is higher than physically reasonable, this pair of values was chosen for convenience to have a 1 [dBm] per gray value step mapping. A smaller (or larger) step size does not have a significant impact on the prediction performance.
}

\tblue{
The gray value $0$ is filled at pixels of building area, which is not our region-of-interest (RoI), while, for our RoI, each pixel is filled with gray values between $1$ and $255$, which corresponds to $P_{\mathrm{RX}}$.
Then, the pathloss map is generated after scaling the considered map scale into a $256\times256$ gray image.
Note that the image size ($256\times256$) has nothing to do with the grayscale ($0-255$).
}

\BfPara{Interpolation}\quad
Since the RT simulations are carried out over a discrete set of RX locations, and it is computationally challenging to gather the channel information for every available RX location, there is missing channel information in a few pixel locations. To fill the missing part of the pathloss map, we utilize \emph{bilinear interpolation}, which approximates the missing value with a weighted sum of the gray values of the adjacent locations.

\BfPara{Data augmentation}\quad
Typically, a larger dataset leads to improved performance of NN training. 
In other words, the larger the data set, the better the outcome. We thus use two augmentation methods - cropping and rotation - to increase the size of our data set. 

The entire map data is cropped into images of about a quarter of the size, taking TX as an anchor point. This augments the size of the dataset by a factor of $96$. 
The image is first cropped as a $64\times64$ size image and then upsampled to a $256\times256$ size image. 
Note that some cropped images, not including any TX, are skipped since the TX location will be used as our second input feature.
After cropping, the image sets are rotated by $90^{\circ}$, $180^{\circ}$, and $270^{\circ}$, thus increasing the size of the dataset by a further factor of $4$.

\BfPara{Sampling}\quad 
In the training and testing of PMNet on the pathloss map dataset, we employ an exclusive division scheme.
Specifically, $90 \%$ and $10 \%$ of images are randomly split into the training and validation set, while the images from the same geographical map belong exclusively to either the training or the validation set. This approach is taken to enhance the generalization performance of PMNet.

\section{Pathloss Map Prediction} \label{sec:pmnet}
\begin{figure*}[t!]
    \centering
    \includegraphics[width=2.\columnwidth]{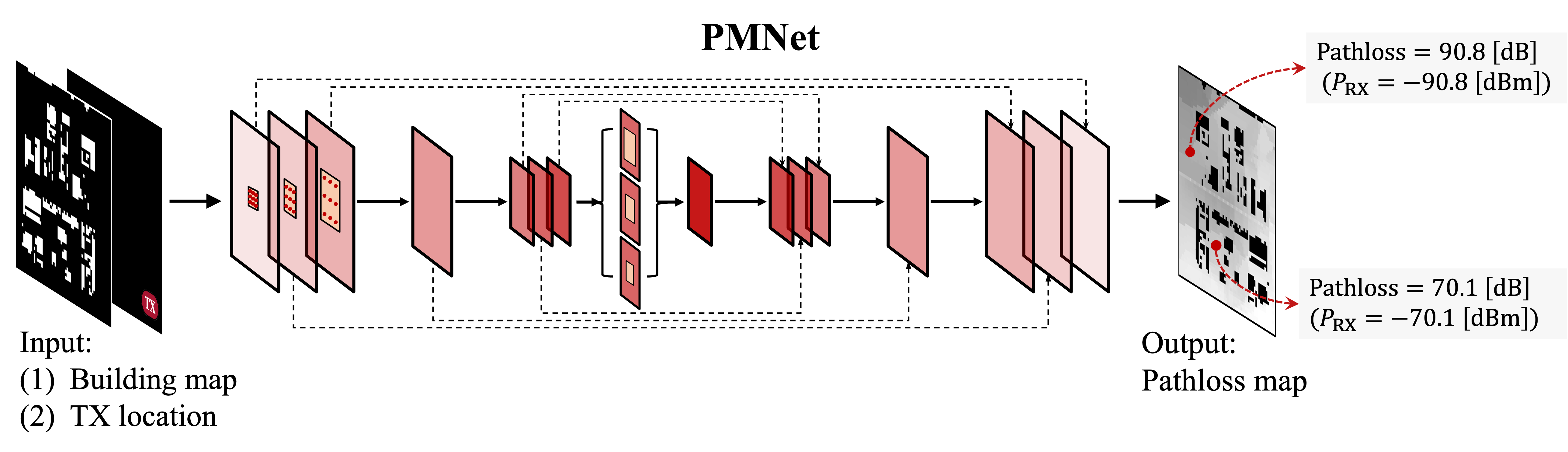}
    \vspace{-.5em}
    \caption{Overview of the PMP task and the PMNet architecture}
    \label{fig:overview_PMNet}
    \vspace{-1.0em}
\end{figure*}

\subsection{Task (1): Pathloss Map Prediction}
We now formulate the prediction task in ML nomenclature. A domain (\textit{i.e.}, wireless channel prediction) is composed of a feature space $\mathcal{X}$, where $x\in{\mathcal{X}}$.
Given the domain, a PMP task is defined as $\mathcal{T} = \{\mathcal{Y}, P(y|\mathbf{x})\}$, which is composed of a label space $\mathcal{Y}$, where $y\in{\mathcal{Y}}$.
Given the task, a dataset is defined as $\mathcal{D} = \{\mathcal{X},\mathcal{Y}\}$, which is a collection of $|\mathcal{D}| 
 = \mathcal{N}$ channel measurement data that belong to a domain with a task $\mathcal{T}$. 

For the PMP task, $\mathcal{X}$ consists of (1) a building map (including terrain, building, and/or foliage) and (2) a TX location and $\mathcal{Y}$ is a Pathloss map.
The goal of the PMP task $\mathcal{T}$ is to find a predictive function $f(\cdot)$, which accurately predicts $\mathcal{Y}$ for a given $\mathcal{X}$.
\tblue{
It is worth noting that integrating RoI (denoted as $\mathcal{A}^{\star}$) segmentation with path gain prediction simplifies the PMP task and eliminates the need for separate pre- or post-processing steps for the RoI segmentation for each map. Additionally, this integration helps NN better understand the different IOs in a given building map.}

In a nutshell, the PMP task is to predict the pathloss/path gain (and received power $P_{\mathrm{RX}}$ using simple normalization) at RX locations $\vb*{q}_{\mathrm{RX}}$ given TX location $\vb*{q}_{\mathrm{TX}}$ in RoI $\mathcal{A}^{\star}$. This channel prediction task exploits \emph{site-specific} geographical information, focusing on the large-scale effects in the channel.



We employ a supervised ML method for the PMP task. We train the model on a dataset of RT channel measurements for an area of $\mathcal{A}$, such as the USC dataset in Table \ref{table:dataset}; see Fig. \ref{fig:overview_PMNet} for an overview of the ML-based PMP approach.


\subsection{Network Architecture}
In this subsection, we present the design process of our proposed PMP-oriented NN architecture, referred to as \emph{PMNet}.
Our design principles are summarized as follows: (1) several state-of-the-art techniques in the field of image processing are carefully selected and tested, (2) some essential techniques are selected following the concept of ablation study, and (3) the NN with selected techniques is optimized with extensive trials.

\begin{table*}[!h]
\small
\resizebox{1.\linewidth}{!}{\begin{minipage}[h]{1.3\linewidth}
\centering
\begin{tabular}{c c c | c c c }
\toprule[1.5pt]
\multicolumn{6}{c}{\textbf{PMNet}} \\
\multicolumn{3}{c}{\textbf{Encoder}} & \multicolumn{3}{c}{\textbf{Decoder}} \\
\cmidrule(lr){1-3} \cmidrule(lr){4-6} 
\textit{\#} & \textit{Type} & \textit{Output Size} & \textit{\#} & \textit{Type} & \textit{Output Size}  \\ 
\midrule[.7pt] 
Input           &  Image  & $2 \times 256 \times 256$        & Output        & Image  &   $1 \times 256 \times 256$ \\ 
1($\downarrow$) &  Conv2d, MaxPool2d    & $64 \times 65 \times 65$   & 1($\uparrow$) & Conv2d & $(128+2) \times 256 \times 256$ \\ 
2               &  ResLayer     &   $256 \times 65 \times 65$   & 2             & Conv2d & $(256+64) \times 65 \times 65$ \\ 
3($\downarrow$) &  ResLayer     & $512 \times 33 \times 33$     & 3             & Conv2d & $(256+256) \times 65 \times 65$ \\ 
4($\downarrow$) &  ResLayer     & $512 \times 17 \times 17$     & 4($\uparrow$) & ConvTranspose2d & $(256+256) \times 65 \times 65$ \\ 
5               &  ResLayer               & $1024 \times 17 \times 17$    & 5($\uparrow$) & ConvTranspose2d & $(512+512) \times 33 \times 33$ \\ 
6               &  Conv2d, AdaptiveAvgPool2d & $512 \times 17 \times 17$ & 6 &  Conv2d & $(512+512) \times 17 \times 17$            \\ 
\midrule[1.5pt]
\end{tabular}

\end{minipage}}
\caption{\tblue{PMNet architectures and parameters. $\downarrow$ and $\uparrow$ represent the downsampling and upsampling layers, respectively.}}
\label{table:pmnet_architecture}
\vspace{-1.em}
\end{table*}

\subsubsection{Design choices}
In the PMP task, the NN is required to perform image segmentation to identify the RoI and predict received power within the RoI, while accounting for complex wireless propagation physics. To accomplish this, our proposed PMNet is designed based on such methods, \emph{Encoder-Decoder} and \emph{Atrous convolution}.

\BfPara{Encoder-Decoder}\quad
Encoder-Decoder networks are a widely applied architecture for many computer vision tasks, \textit{\textit{e.g.}}, object detection \cite{araki2020deconvolutional}, human pose estimation \cite{hourglass}, and semantic segmentation \cite{Deeplabv3+,lin2017multi,peng2017large}.
The encoder-decoder architecture allows to learn a lower-dimensional representation from a higher-dimensional dataset and utilize the learned representation for various tasks. However, as the encoder shrinks the input feature maps, it may lose essential information, leading to a \emph{bottleneck problem}. Several architectures, including U-Net \cite{UNet}, address the bottleneck problem by adding \emph{skip connections} between the encoder and the decoder parts. Skip connections allow the decoder to access feature maps from the encoder, which helps to propagate context information to higher-resolution layers.

\BfPara{Atrous convolution}\quad
\emph{Receptive field} of a convolutional layer is the region of the input feature map that contributes to the output feature map at a given location. The size of the receptive field is determined by the resolution of the input feature map and the field-of-view (FoV) of the filter.
There is a logarithmic relationship between the localization accuracy of a model and the size of its receptive field. This means the receptive field size should be sufficient if the given dataset and task are observed with wide FoV.
A standard convolutional filter detects a particular feature by sliding over the input feature map, resulting in the output feature map seeing only the adjacent part of the input feature map. In terms of computational complexity, having a wide receptive field with the standard convolutional filter is expensive. Thus, broadly speaking, the receptive field of the standard convolution filter is somewhat narrow, seeing only little context.

Atrous convolution, also known as dilated convolution, is a technique that addresses this limitation \cite{fisher2016multi}. It allows capturing a larger context with a wider FoV by modifying the standard convolution operation. For the two-dimensional case, atrous convolution is applied over the input feature map $f$ to produce the output feature map $g$ at location $\{i, j\}$ using the convolution filter $w$. This operation can be expressed as follows:
\begin{equation}
    g_{\{i, j\}} = \sum_{m=1}^{k}\sum_{n=1}^{k} f_{ \{i+ rm, j + rn\} }  w_{\{m, n\}}. \label{eq:AstrousConvolution}
\end{equation}
Here, $k$ represents the kernel size, and $r$ is the atrous rate, which determines the stride level. Notably, the atrous rate $r$ allows to adaptively control the FoV of the filter. For example, an atrous rate of $r=2$ doubles the FoV of the filter, while an atrous rate of $r=3$ triples it. The standard convolution can be seen as a special case of \eqref{eq:AstrousConvolution} where $r = 1$.

\tblue{
In the context of the PMP task, the encoder-decoder symmetric architecture of PMNet facilitates efficient context propagation from the encoder to the decoder, while atrous convolution enables it to handle scale variations and capture broader context in map data, setting it apart from other UNet-based networks \cite{RadioUNet, Fadenet, enes2023agile, kehai2023deep}.
The combination of these two features enables PMNet to efficiently and accurately predict pathloss maps, while also accounting for complex wireless propagation physics.
}

\subsubsection{Design parameters}
PMNet architectures are composed of a stack of \emph{ResLayers}, each containing multiple residual blocks \cite{reslayer}. These ResLayers can be configured with varying numbers of blocks, atrous rates, multi-grids, and output strides. These elements are summarized as follows:
\begin{itemize}
    \item \emph{Number of blocks}: The number of residual blocks in a ResLayer controls the complexity and depth of the network. Increasing the number of blocks may improve the accuracy of the model, but it also increases the computational cost.
    \item \emph{Atrous rates}: Atrous rates control the spacing between the convolutions in a ResLayer. Larger atrous rates allow the network to capture more larger spatial contexts in the PMP task.
    \item \emph{Multi-grids}: Multi-grids allow the network to capture multi-scale information from different levels of the CNN architecture. 
    \item \emph{Output stride}: The output stride of a ResLayer controls the ratio between the resolution of the input image and the output image's resolution. A higher output stride results in a lower-resolution output image. This can be useful to strike a balance between accuracy and speed.
\end{itemize}
Note that the impact of output stride in the PMP task is shown in Table \ref{table:ablation} in Sec. \ref{sec:pmnet_simulation} (\textit{e.g.}, the case of $\frac{H}{8}\times\frac{W}{8}$).
With these design choices and parameters, PMNet effectively predicts pathloss maps even for different channel measurement datasets (\textit{e.g.}, \emph{RadioMapSeer} \cite{RadioMapSeer}).
For an architectural overview, please refer to Fig. \ref{fig:overview_PMNet} and Table \ref{table:pmnet_architecture}. For more details, please see our source code repository.

\subsection{Training} 
Table~\ref{table:parameter_train} lists the hyper-parameters that are used for the training of PMNet. We implement the PMNet using PyTorch and use an NVIDIA GeForce RTX 3080 Ti GPU. For more stable training, we normalize the input values into $[0, 1]$ via scaling. During the training, we evaluate the PMNet by mean squared error (MSE) on the validation set at the end of every epoch. For testing, we use the parameters of PMNet with the best MSE score on the validation set. Consequently, the pathloss map for a given map can be generated within a few milliseconds after training.

\begin{table}[!h]
\centering
\small
\begin{tabular} {l l}
\toprule[1.5pt]
\textit{\textbf{Model}} & PMNet \\
\midrule[.7pt] 
\textit{\textbf{Dataset}} (USC) &  \\
\cmidrule(lr){1-1} \cmidrule(lr){2-2}
Map & USC campus \\
Split for training (test) set & $90\%$ ($10\%$) of dataset \\ 
\midrule[.7pt] 
\textit{\textbf{Hyper-parameter}} & \textit{\textbf{}} \\
\cmidrule(lr){1-1} \cmidrule(lr){2-2}
Learning rate (LR) & $10^{-3}\sim 5\times10^{-4}$  \\ 
LR gamma, step size & $0.5$, $10$ \\
Batch size & $16\sim32$ \\ 
Optimizer & Adam \\
\# of of epochs & $50$ \\
\bottomrule[1.5pt]
\end{tabular}

\caption{Training configuration and hyper-parameters for PMNet training.}
\label{table:parameter_train}
\vspace{-1.em}
\end{table}

\begin{table*}[h!t]
\centering
\resizebox{.85\linewidth}{!}{\begin{minipage}{.8\linewidth}
\centering
\begin{tabularx}{1\linewidth}{l||c c ||c c c}
\toprule[1.5pt]
\textit{\textbf{Case}} & \textit{\textbf{Data Aug. ($\times 4$)}} & \textit{\textbf{Feature Size}} & \textit{\textbf{RMSE}$\downarrow$} & \textit{\textbf{RoI Segmentation Err.}$\downarrow$} & \textit{\textbf{Channel Prediction Err.}$\downarrow$}  \\
\cmidrule(lr){1-1} \cmidrule(lr){2-2} \cmidrule(lr){3-3} \cmidrule(lr){4-4} \cmidrule(lr){5-5}  \cmidrule(lr){6-6}

w/o Data-Aug.  & \xmark & $\frac{H}{16}\times\frac{W}{16}$ 
& 0.01637 & 0.00263 & 0.01860
\\
w/ Data-Aug.   & \cmark & $\frac{H}{16}\times\frac{W}{16}$
& 0.01259 & 0.00025 & 0.01403
\\
$\frac{H}{8}\times\frac{W}{8}$  & \cmark & $\frac{H}{8}\times\frac{W}{8}$
& 0.01057  & 0.00096 & 0.01175
\\
\bottomrule[1.5pt]
\end{tabularx}
\end{minipage}}
\caption{Ablation study for PMNet training optimization. Lower values indicate better performance.}
\label{table:ablation}
\vspace{-1.em}
\end{table*}

\subsection{Evaluation}

\BfPara{Root mean square error (RMSE)}\quad
RMSE is a widely used loss function in regression analysis and is used as the primary evaluation metric for this task. It measures the overall difference between the prediction $\hat{\vb* y}$ and ground-truth $\vb* y$ and quantifies the overall accuracy of the model. The formula for RMSE is:
\begin{equation}
\mathrm{RMSE}(\hat{\vb* y}, \vb* y) = \sqrt {\frac{1}{N} \sum_{n=1}^{N} (\hat{y}_{n} - y_{n})^2},
\end{equation}
where $\hat{y}_{n} \in \hat{\vb* y}$ and $y_{n} \in \vb* y$ denote predicted and ground-truth gray value (corresponding $P_{\mathrm{RX}}$) at the $n$-th pixel, respectively, and $N$ is the number of pixels in a pathloss map, \textit{i.e.}, $256 \times 256$. 
The RMSE averaged over all samples is the primary evaluation metric for the PMP task.

\BfPara{\tblue{RoI segmentation error}}\quad
\tblue{
The RoI segmentation error, calculated using the intersection over union (IoU) metric, quantifies the accuracy of RoI and non-RoI area segmentation for all pixels in the ground-truth (${\{i,j\}}$) and prediction ($\{\hat{i},\hat{j}\}$) - that is calculated as follows:}
\begin{align}
\mathrm{RoI~Segmentation~Err.} = \frac{\sum_{i}\sum_{j} \mathrm{Err^{B}}_{\{i,j\}} } {\sum_{i}\sum_{j} \mathrm{Bld}_{\{i,j\}}}.
\end{align}
Here, $\mathrm{Err^{B}}_{\{i,j\}}$ and $\mathrm{Bld}_{\{i,j\}}$ are defined as:
\begin{align}
 \mathrm{Err^{B}}_{\{i,j\}} &= 
\begin{cases}
    1, \ \{i,j\} \in \mathcal{B}~\mathrm{and}~\{\hat{i},\hat{j}\} \in \mathcal{A}^{\star} \\ 
    1, \ \{i,j\} \in \mathcal{A}^{\star}~\mathrm{and}~\{\hat{i},\hat{j}\} \in \mathcal{B} \\ 
    0, \ \mathrm{otherwise}
\end{cases} \\
 \mathrm{Bld}_{\{i,j\}} &= 
\begin{cases}
    1, \ \{i,j\} \in \mathcal{B} \\ 
    0. \ \mathrm{otherwise}
\end{cases} 
\end{align}
\tblue{Within a given map, the non-RoI area, denoted as black (gray value $0$), is represented by $\mathcal{B}$, while the RoI area, denoted as non-black (grayscale $1-255$), is represented by $\mathcal{A}^{\star}$. $\mathcal{B}$ and $\mathcal{A}^{\star}$ are complementary set within $\mathcal{A}$.
$\mathcal{B}$ can include buildings, foliage, and/or small objects.}

\BfPara{Channel prediction error}\quad
\tblue{
Channel prediction error directly evaluates path gain accuracy for pixels within the RoI area, evaluating power in [dBm] (or path gain in [dB]) unlike RMSE, which quantifies differences based on gray values.

To calculate channel prediction error, gray values within the RoI area of both the predicted and ground-truth pathloss maps are converted into corresponding received power values. The RMSE formula is then applied to these power values}:
\begin{equation}
\mathrm{RMSE}(\hat{\vb* p}, \vb* p) = \sqrt {\frac{1}{N} \sum_{n=1}^{N} (\hat{p}_{n} - p_{n})^2},
\end{equation}
where $\hat{p}_{n} \in \hat{\vb* p}$ and $p_{n} \in \vb* p$ represent the predicted and ground-truth  $P_{\mathrm{RX}}$ at the $n$-th pixel, respectively.
Channel Prediction Error is then computed by averaging $\mathrm{RMSE}(\hat{\vb* p}, \vb* p)$ across all given samples.



\subsection{Simulation Result} \label{sec:pmnet_simulation}

\subsubsection{Training optimization}
Table \ref{table:ablation} presents an ablation study to identify the factors that significantly contribute to PMNet's performance in the PMP task, such as \emph{data augmentation} and \emph{feature map size}.\footnote{\tblue{Our extensive experiments tested other factors, such as different sampling methods, training loss functions, and additional input features (\textit{e.g.}, TX distance heatmap), but these factors did not show a meaningful improvement to justify the additional complexity.}}

\BfPara{Impact of data augmentation}\quad
For the data augmentation, we do horizontal, vertical and diagonal flips. In other words, including the original images, we use the $\times 4$ number of images for training. Note that data augmentation has several advantages in general: first, it enhances the diversity of the training data by generating additional examples that capture various variations of the original data. Second, it reduces overfitting by exposing the model to a wider range of input patterns. Finally, data augmentation helps to make the model more robust to noise and variability in the input data. As shown in Table~\ref{table:ablation}, it improves the performance of PMNet by $15.7$\% in terms of RMSE.

\BfPara{Impact of feature map size}\quad
We analyze the performances of PMNet according to the size of the feature map, which is the output of the encoder. Table~\ref{table:ablation} compares the results with the feature map sizes $\frac{H}{8} \times \frac{W}{8}$ and $\frac{H}{16} \times \frac{W}{16}$, where $H$ and $W$ are the height and width of an input image, respectively. To adjust the feature map size, we modify the strides of the convolution layers in the encoder. We employ the feature map size of $\frac{H}{8} \times \frac{W}{8}$ as the default option, because PMNet yields better performance with the feature map size of $\frac{H}{8} \times \frac{W}{8}$ than that of $\frac{H}{16} \times \frac{W}{16}$. 

\begin{figure*}[h!]
\centering
\begin{subfigure}[t]{.23\textwidth}
\centering
\includegraphics[width=\linewidth]{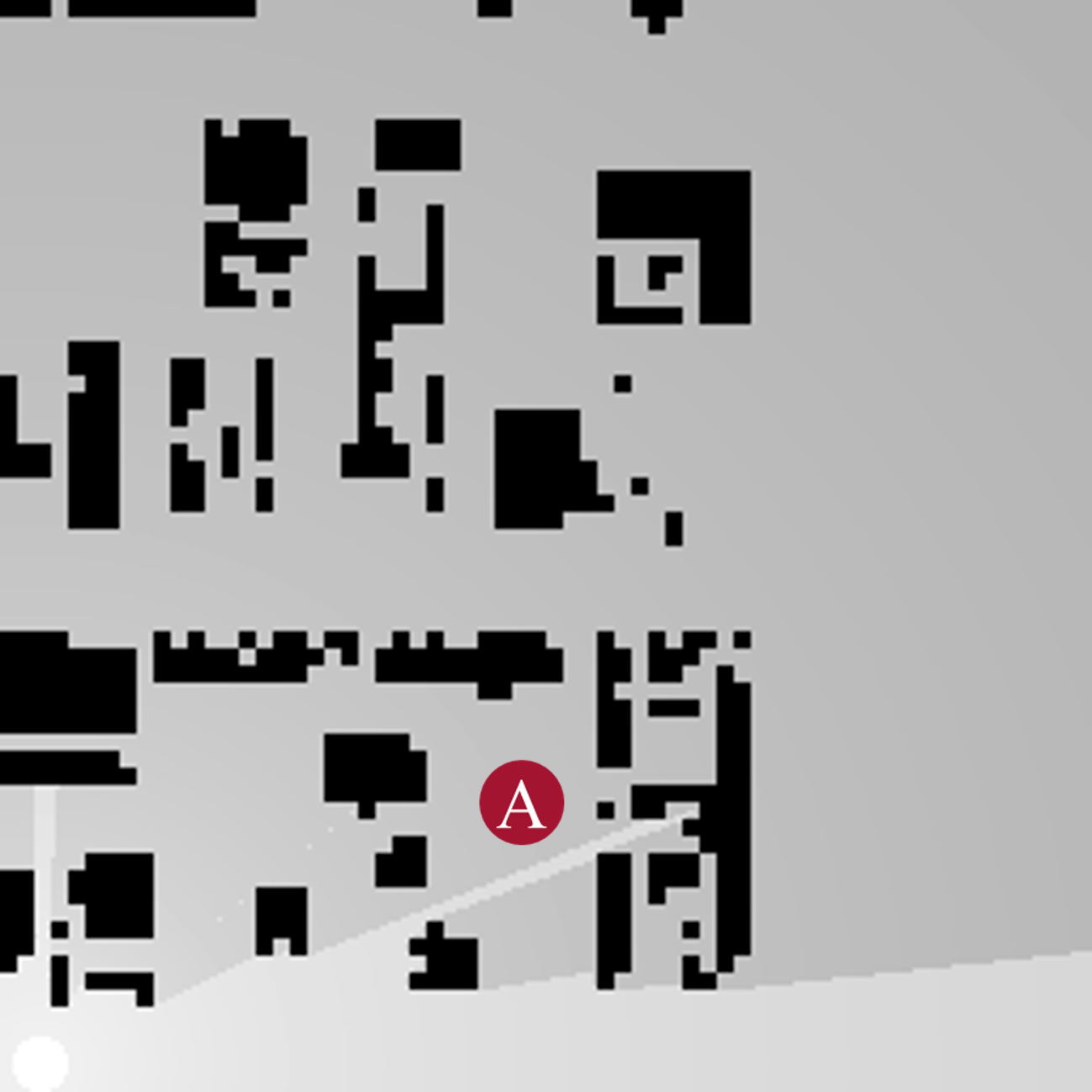}
\caption{Prediction (3GPP)}
\label{comparison:3GPP}
\end{subfigure}\hspace*{.025\textwidth}%
\begin{subfigure}[t]{.23\textwidth}
\includegraphics[width=\linewidth]{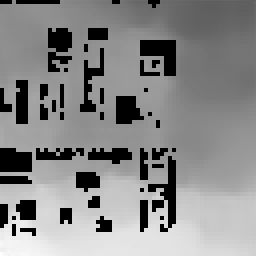}
\caption{Prediction (RadioUNet)}
\label{comparison:RadioUNet}
\end{subfigure}\hspace*{.025\textwidth}%
\begin{subfigure}[t]{.23\textwidth}
\centering
\includegraphics[width=\linewidth]{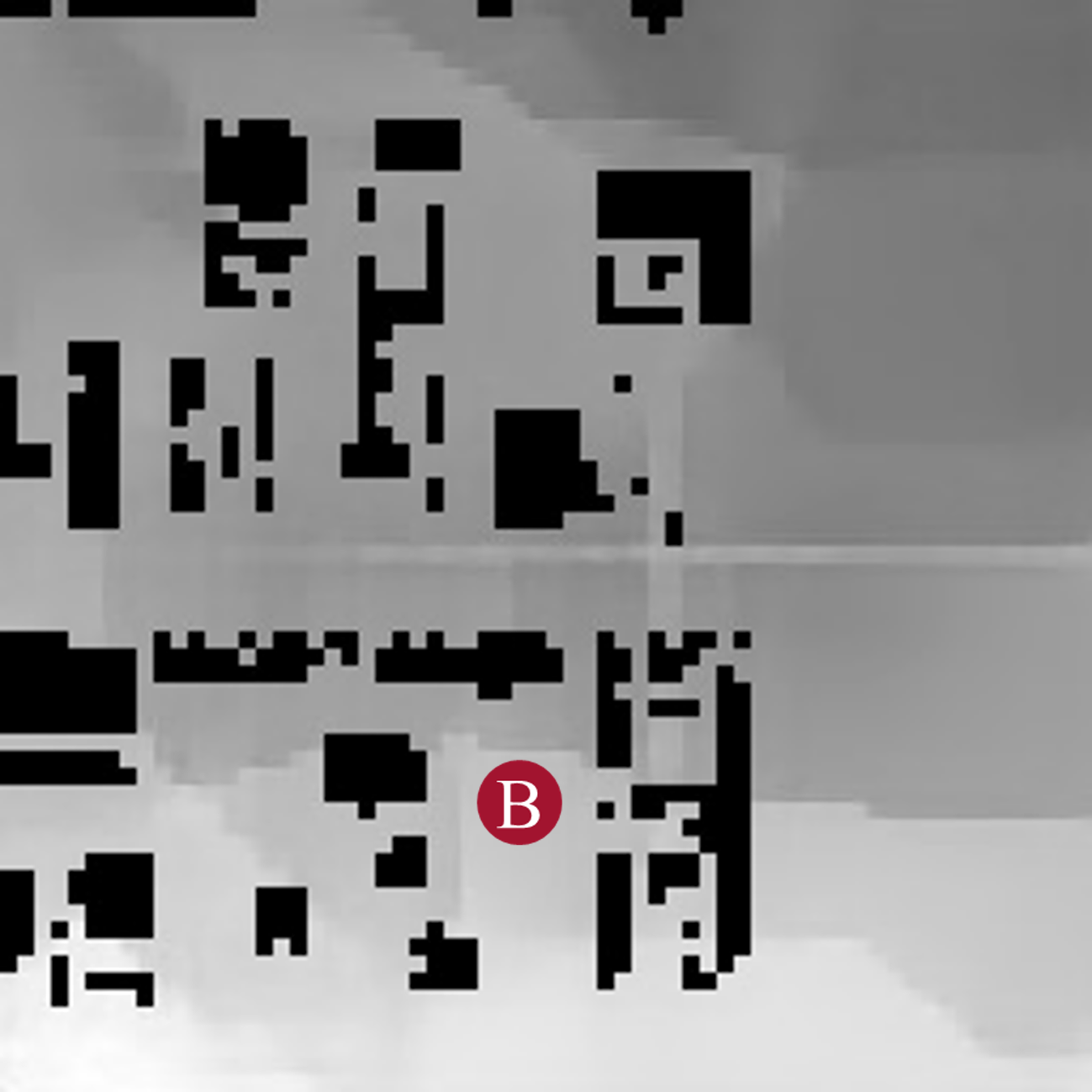}
\caption{Prediction (PMNet)}
\label{comparison:PMNet}
\end{subfigure}\hspace*{.025\textwidth}%
\begin{subfigure}[t]{.23\textwidth}
\centering
\includegraphics[width=\linewidth]{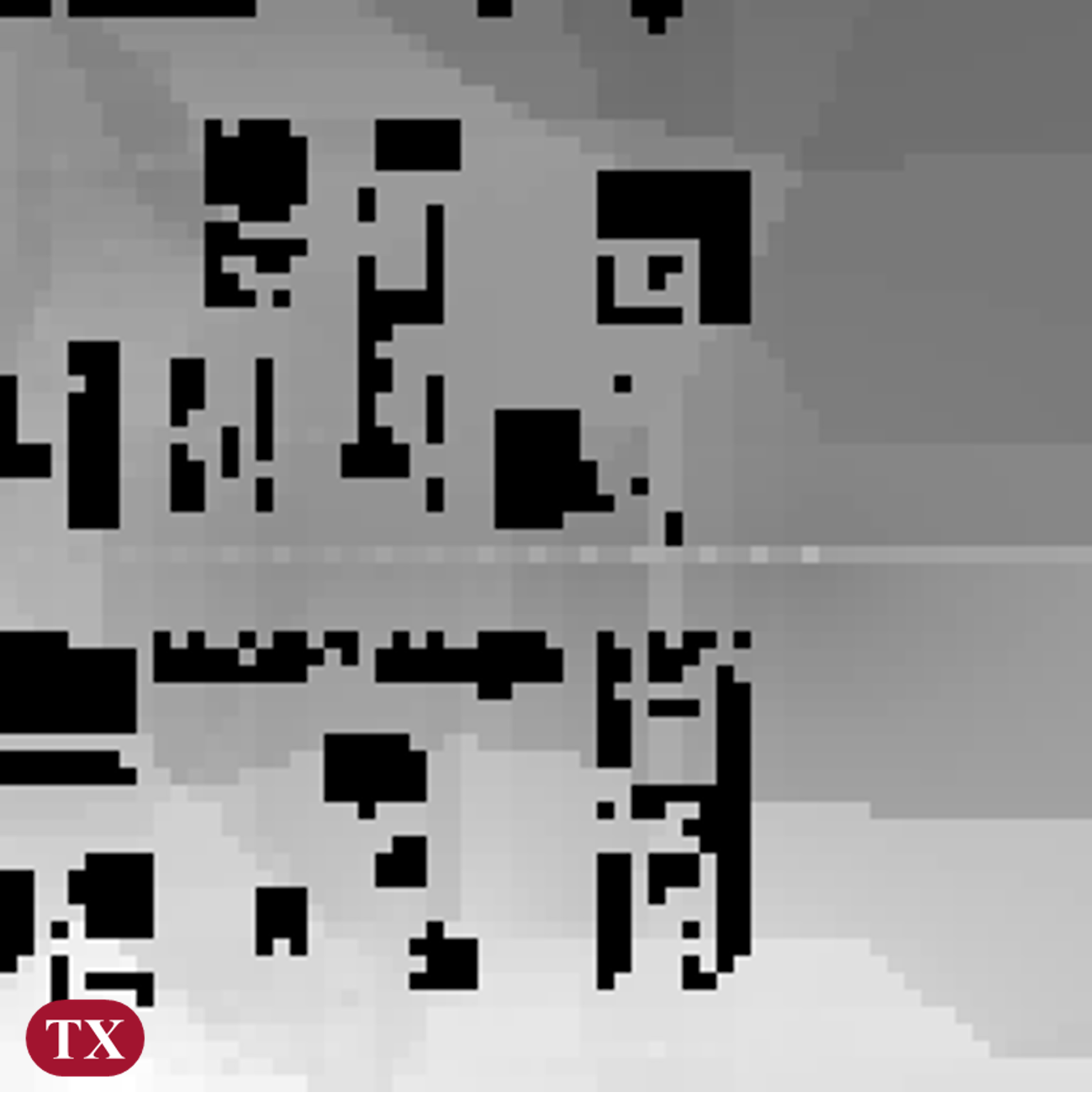}
\caption{Ground-truth (\emph{Wireless Insite})}
\label{comparison:GT}
\end{subfigure}\hspace*{.025\textwidth}
\\ \vspace*{.01\textwidth}
\begin{subfigure}[t]{.23\textwidth}
\centering
\includegraphics[width=\linewidth]{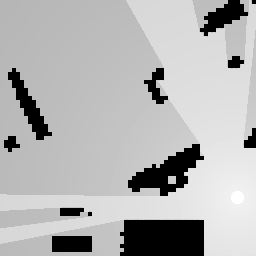}
\caption{Prediction (3GPP)}
\label{comparison:3GPP3}
\end{subfigure}\hspace*{.025\textwidth}%
\begin{subfigure}[t]{.23\textwidth}
\includegraphics[width=\linewidth]{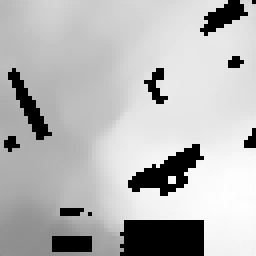}
\caption{Prediction (RadioUNet)}
\label{comparison:RadioUNet3}
\end{subfigure}\hspace*{.025\textwidth}%
\begin{subfigure}[t]{.23\textwidth}
\centering
\includegraphics[width=\linewidth]{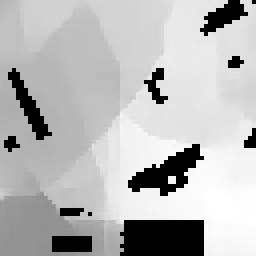}
\caption{Prediction (PMNet)}
\label{comparison:PMNet3}
\end{subfigure}\hspace*{.025\textwidth}%
\begin{subfigure}[t]{.23\textwidth}
\centering
\includegraphics[width=\linewidth]{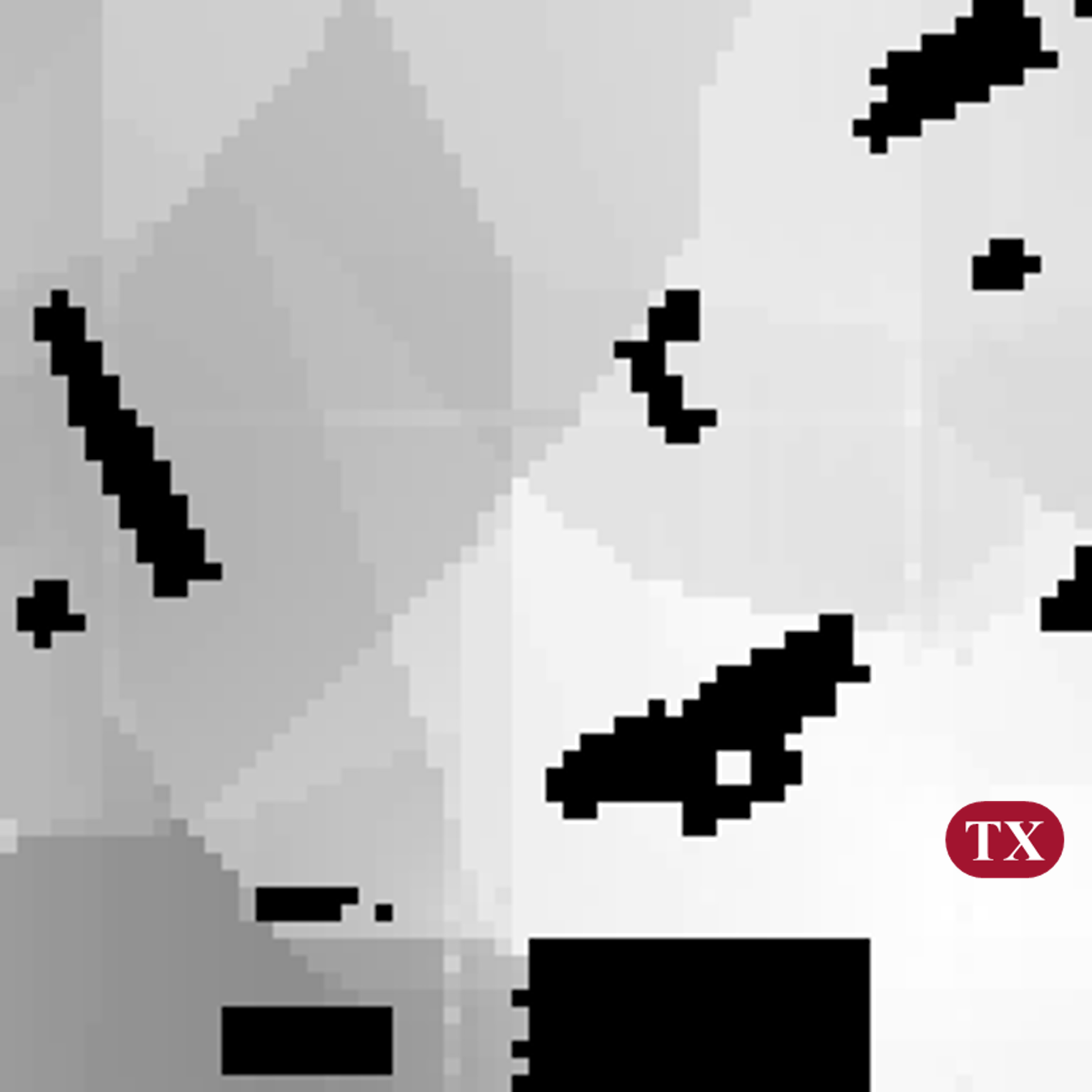}
\caption{Ground-truth (\emph{Wireless Insite})}
\label{comparison:GT3}
\end{subfigure}\hspace*{.025\textwidth}
\caption{\tblue{Comparison of the predicted pathloss map of 3GPP, RadioUNet, and PMNet. \tinycircled{\textbf{TX}} in ground-truth represents the TX location. The scenes are randomly selected, not cherry-picked.}}
\label{fig:comparison}
\end{figure*}
\begin{table*}[h!]
\centering          
\resizebox{0.83\linewidth}{!}{\begin{minipage}[h]{.74\linewidth}
\centering

\newcolumntype{R}{>{\raggedleft\arraybackslash}X}
\begin{tabularx}{1\linewidth}{l || c || c c c}
\toprule[1.5pt]
\textit{\textbf{Scheme}} & \textit{\textbf{ML-based}} & \textit{\textbf{RMSE}}$\downarrow$ & \textit{\textbf{RoI Segmentation Err.}}$\downarrow$ & \textit{\textbf{Channel Prediction Err.}}$\downarrow$  \\
\cmidrule(lr){1-1} \cmidrule(lr){2-2} \cmidrule(lr){3-3} \cmidrule(lr){4-4} \cmidrule(lr){5-5} 
3GPP (with map info.) \cite{3gpp2018pathloss} & \xmark & 15.9451 & - & 17.5973
 \\
RadioUNet \cite{RadioUNet} & \cmark & 0.02634 & 0.00840 & 0.01249 \\
PMNet & \cmark & \textbf{0.01057}  & \textbf{0.00096} & \textbf{0.01175} \\
\bottomrule[1.5pt]
\end{tabularx}

\centering
\end{minipage}}
\caption{Comparison study for PMP schemes (3GPP, RadioUNet, and PMNet). Lower values indicate better performance, and the lowest errors are highlighted.}
\label{table:comparison_baseline}
\vspace{-1.em}
\end{table*}

\subsubsection{Accuracy} \label{sec:comparison sutdy}
We compare the ML-based PMP with our proposed PMNet model to two other methods for the PMP task: a model-based approach, \emph{3GPP}, and an ML-based approach, \emph{RadioUNet}. All three methods produce a single-channel $256 \times 256$ image of the pathloss map as the output, given the input of a two-channel $256 \times 256$ image containing the geographical map and the TX location. Here are the details of these baseline methods:
\begin{enumerate}
\item \textbf{3GPP (with map info.)} As discussed in Sec. \ref{sec:3gpp model}, the 3GPP model determines the pathloss at a particular location based on the Euclidean distance and whether the link between the TX and RX is in LoS or NLoS.  
To ensure a fair comparison with other baselines, we utilize map information to determine the LoS or NLoS condition of specific pixels to the TX.\footnote{
The original 3GPP pathloss model uses a probabilistic model to determine LoS/NLoS condition at a particular distance. However, to ensure a fair comparison, we use here the deterministic LoS/NLoS condition determined from the map information in calculating the pathloss gain.
} Note that it does not require any NN training as it is a model-based approach.

\item \textbf{RadioUNet} \cite{RadioUNet} is an ML-based PMP method that extends the UNet architecture by employing two UNets. Each UNet comprises $8$ encoder layers with convolution, ReLU, and Maxpool layers, followed by $8$ decoder layers with transposed convolution and ReLU layers. The encoders and decoders are concatenated, as in the original UNet architecture. Here, RadioUNet employs curriculum training to enhance training: in the first stage, the first UNet is trained for a specific number of epochs, with the second UNet frozen. In the second stage, the second UNet is trained using the two-channel input features and the output of the first UNet, effectively making it a three-channel input network.
\item \textbf{PMNet (Proposed)} is our proposed ML-based PMP method.
This network employs several parallel atrous convolutions with different rates and the encoder-decoder network. 
The encoder consists of $6$ ResNet-based layers. Each ResNet layer comprises several bottleneck layers consisting of convolution, batch normalization, max pooling, and ReLU. 
The decoder consists of $6$ layers consisting of convolution, adaptive average pooling, ReLU, transposed convolution, and ReLU. 
Skip connections are used between encoders and decoders.
\end{enumerate}

\BfPara{Qualitative analysis}\quad
Fig. \ref{fig:comparison} shows the prediction results of the baselines. Recall that each pixel in the RoI corresponds to the predicted received power $P_{\mathrm{RX}}$ (or the path gain $\mathrm{PG}$).
Note that some pixel values in the ground-truth data appear noisy due to interpolation during the gray conversion process after RT simulation.

\tblue{
3GPP exhibits a substantial deviation from ground-truth obtained through RT simulation, highlighting the differences between how RT simulation and 3GPP model calculate a pathloss.
\tred{
Specifically, for RX locations with LoS conditions close to the TX, the results obtained using the 3GPP model approximately match the ground-truth data obtained from \emph{Wireless Insite}. However, for RX locations farther from the TX or under non-LoS conditions, the 3GPP model exhibits significant discrepancy from the ground-truth data.}
It is worth noting that the 3GPP pathloss model does not provide results for near-field within a link distance of 10 meters; so, we arbitrarily set the power in the near-field area to gray value $255$, which does not introduce significant errors.
The 3GPP pathloss model is a simplified model that does not account for the complex wireless propagation physics of reflection, diffraction, and scattering (highlighted in \circled{A}). Instead, it relies solely on two models for LoS and NLoS locations, respectively, and only considers link distance and carrier frequency. This simplified approach inevitably leads to significant inaccuracies in the pathloss prediction.

RadioUNet demonstrates impressive RoI segmentation results, while its channel prediction outputs appear somewhat blurry. It is worth noting that RadioUNet conducts curriculum-based training with 50 epochs each in the first and second stages, utilizing the same training/validation set as PMNet, which is trained with a total of 50 epochs. 

PMNet, on the other hand, achieves notable results for both RoI segmentation and channel prediction. 
As highlighted in \circled{B}, PMNet effectively captures the intricate wireless propagation physics of reflection, diffraction, and scattering. This can be attributed to PMNet's ability to incorporate a broader contextual understanding of the environment, enabling it to capture the representation of wireless propagation physics in the surrounding environment.}


\begin{table*}[h!]
\centering          
\resizebox{1.\linewidth}{!}{\begin{minipage}[h]{.93\linewidth}
\centering        
\begin{tabular}{l|| c c c || c c c }
\toprule[1.5pt]
\textit{\textbf{Case}} & \textit{\textbf{Model}} & \textit{\textbf{Train Data}} & \textit{\textbf{Eval. Data}} & \textit{\textbf{RMSE}}$\downarrow$ & \textit{\textbf{RoI Segmentation Err.}}$\downarrow$ & \textit{\textbf{Channel Prediction Err.}}$\downarrow$  \\
\cmidrule(lr){1-1} \cmidrule(lr){2-2} \cmidrule(lr){3-3} \cmidrule(lr){4-4} \cmidrule(lr){5-5} \cmidrule(lr){6-6}  \cmidrule(lr){7-7} 
Vanilla & PMNet & USC & USC  & 0.01057 & 0.00096 & 0.01175 \\
\midrule[.1pt]
Cross-scenario (UCLA) & PMNet & USC & UCLA   & 0.19146
& 0.03925 & 0.21700 \\
Cross-scenario (Boston) & PMNet & USC & Boston & 0.25842
& 0.04602 & 0.32436 \\
\bottomrule[1.5pt]
\end{tabular}


\end{minipage}}
\caption{Numerical results of PMNet on an unseen network scenario. PMNet was trained on the USC dataset and evaluated on the UCLA and Boston dataset.}
\label{table:zero-shot}
\end{table*}

\BfPara{Quantitative analysis}\quad
Table \ref{table:comparison_baseline} compares our proposed PMNet model to the model-based 3GPP method and the ML-based RadioUNet method in terms of three accuracy metrics for the PMP task: RMSE, RoI segmentation error, and channel prediction error. Note that the ground-truth dataset is made by RT simulation; therefore, the error shows the difference between a scheme and the measurement by RT simulation.

The model-based 3GPP method has inferior results compared to ML-based methods, which can be explained by the oversimplifications inherent in this model, as discussed above. While our proposed PMNet model achieves the best score on all three metrics, another ML-based PMP method, RadioUNet, also achieves high accuracy ($RMSE~\leq 0.03$).
This result highlights the capability of ML-based PMP approaches to learn a representation of the wireless propagation physics implicit in the ground-truth RT measurement data.



\section{Transferable Pathloss Map Prediction} \label{sec:transferlearning}
\subsection{Challenge: PMP for Unseen Network Scenario}
\begin{figure}[h!]
\centering
\begin{subfigure}[t]{.48\linewidth}
\centering
\includegraphics[width=\linewidth]{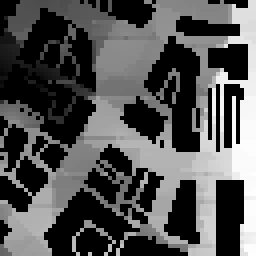}
\caption{Ground-truth (\emph{Wireless Insite})}
\end{subfigure}\hspace*{.015\textwidth}%
\begin{subfigure}[t]{.48\linewidth}
\centering
\includegraphics[width=\linewidth]{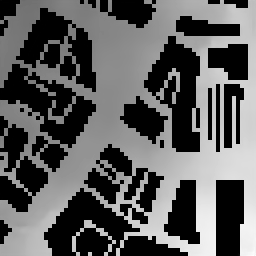}
\caption{Prediction (PMNet)}
\end{subfigure}\hspace*{.015\textwidth}%
\caption{Prediction results of PMNet on an unseen network scenario (\textit{i.e.}, cross-scenario evaluation). The model is trained on the USC dataset and evaluated on the Boston dataset.}
\label{fig:comparison_zeroshot}
\end{figure}

\tblue{
As demonstrated in the previous section, PMNet exhibits high accuracy of the PMP task for a given dataset. 
However, minimizing re-training efforts for new network scenarios remains a challenge.
To evaluate PMNet's generalizability across different scenarios, we conducted a cross-scenario evaluation, testing PMNet trained on USC data on the Boston dataset.

As shown in Fig. \ref{fig:comparison_zeroshot} and Table \ref{table:zero-shot}, the PMNet achieves the RoI segmentation error on the order of $10^{-2}$ and the channel prediction error on the order of $10^{-1}$ in a new scenario. Such deterioration is due to differences in network configuration and environmental characteristics between the two scenarios (\textit{e.g.}, different map scales and geographical features).
This highlights the need for further development to improve PMNet's performance across different network scenarios, a task we refer to as \emph{cross-scenario PMP}.
}


\subsection{Task (2): Cross-scenario PMP}
To enable better performance, we now allow cross-scenario PMP to {\em improve} the model trained on a different network scenario through training with a {\em reduced-size} training in the new scenario. This will allow the network to adapt to the new scenario with less time and resource effort, while maintaining high accuracy.
To address this challenge, we leverage \emph{transfer learning} (TL).

\BfPara{Approach: Transfer learning}\quad
TL is an ML technique that allows knowledge transfer from one task or dataset to another. In the context of cross-scenario PMP, we can transfer the knowledge from the source scenario, which learns a predictive function $f_S(\cdot)$ from a source dataset $\mathcal{D}_S$ (\textit{e.g.}, USC), to the target scenario, which learns a predictive function $f_T(\cdot)$ from a target dataset $\mathcal{D}_T$ (\textit{e.g.}, UCLA and Boston). 

There are two main ways to use TL for the cross-scenario PMP. 
\begin{itemize}
    \item \emph{Feature extraction}: We can train a feature extractor on a source scenario and then use that feature extractor to extract features from data from a target scenario. Once we have extracted the features, we can train a simple model (\textit{e.g.}, a linear regressor) to predict the pathloss map for the target scenario.
    \item \emph{Fine-tuning}: We can fine-tune a pre-trained model on the target scenario. This can be done by unfreezing some or all of the layers of the pre-trained model and training the model on data from the target scenario.
\end{itemize}
The choice between those two methods depends on a number of factors, including the size and complexity of the pre-trained model, the availability of training data for the target dataset, and the computational resources available.

In this work, we focus on the fine-tuning TL approach with all of the layers of the pre-trained model unfrozen.\footnote{\tred{While we have performed sample experiments with unfreezing certain layers, such as the encoder-frozen and decoder-unfrozen, performance did not improve significantly. A more comprehensive investigation of this topic is, however, beyond the scope of this paper.}} This approach is simple yet effective, achieving higher accuracy on various cross-scenario PMP tasks with less training data and shorter training time, as elaborated in the following subsection.


We prepare and use the following pre-trained models in our experiments:
\begin{enumerate}[label=(\roman*)]
    \item \textbf{VGG16$_{\mathrm{\mathbf{ImgNet}}}$} is the pre-trained CNN model trained on the ImageNet dataset, which contains 140k images belonging to 22k categories. It is a powerful image classification model that has been used to achieve state-of-the-art results on a variety of image classification benchmarks.
    \item \textbf{PMNet$_{\mathrm{\mathbf{3gpp}}}$} is the pre-trained PMNet model trained on the 3GPP pathloss map dataset. The 3GPP pathloss map dataset is prepared with the 3GPP pathloss model in \cite{3gpp2018pathloss} (see 3GPP in Sec. \ref{sec:comparison sutdy}, Fig. \ref{fig:comparison}, and Table \ref{table:comparison_baseline}).
    \item \textbf{PMNet$_{\mathrm{\mathbf{usc}}}$} is the pre-trained PMNet model trained on the USC RT dataset. It is similar to PMNet$_{\mathrm{\mathbf{3gpp}}}$ but is trained on a different dataset. This is our main pre-trained model.
\end{enumerate}
Each pre-trained model is available on our GitHub page.

\begin{table}[!h]
\centering
\small
\resizebox{1.\linewidth}{!}{\begin{minipage}[h]{1.09\linewidth}
\begin{tabular} {l l}
\toprule[1.5pt]
\textit{\textbf{Model}} & \textit{\textbf{}} \\
\cmidrule(lr){1-1} \cmidrule(lr){2-2}
Backbone & PMNet, VGG16 \\
Pre-trained model & PMNet$_{\mathrm{usc}}$, PMNet$_{\mathrm{3gpp}}$, VGG16$_{\mathrm{{ImgNet}}}$ \\
\midrule[.7pt] 
\textit{\textbf{Dataset}} (UCLA, Boston) & \textit{\textbf{}} \\
\cmidrule(lr){1-1} \cmidrule(lr){2-2} 
Map & UCLA campus, Boston \\
Split for training (test) set & $10\%\sim90\%$ ($10\%$) of dataset \\ 
\midrule[.7pt] 
\textit{\textbf{Hyper-parameter}} & \textit{\textbf{}} \\
\cmidrule(lr){1-1} \cmidrule(lr){2-2}
LR & $10^{-3}\sim5\times10^{-4}$  \\ 
LR gamma, step size & $0.5$, $10$ \\
Batch size & $16$ \\ 
Optimizer & Adam \\
\# of of epochs & $50$ \\
\bottomrule[1.5pt]
\end{tabular}

\end{minipage}}
\caption{\tblue{Training configuration and hyper-parameters in cross-scenario PMP.}}
\label{table:parameter_network_finetuning}
\vspace{-1.em}
\end{table}

\begin{figure*}[h!]
    \centering
    \includegraphics[width=1.85\columnwidth]{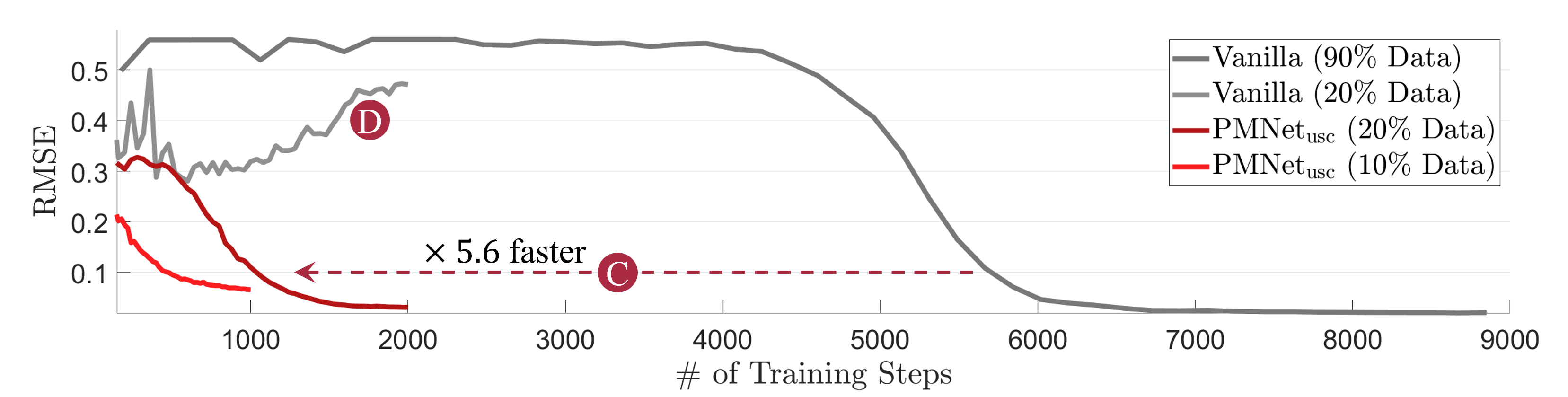}
    \caption{\tblue{Comparison of the training efficiency of PMNet models with and without TL. PMNet models are trained for 50 epochs and evaluated on the Boston dataset.}}
    \label{fig:comparisn_TL_val}
    \vspace{-0.5em}
\end{figure*}

\subsection{Simulation Results}

\tblue{
As demonstrated in the cross-scenario evaluation results (in Fig. \ref{fig:comparison_zeroshot} and Table \ref{table:zero-shot}), there is a need for further development to make PMNet adapt to different network scenarios. To this end, our approach is fine-tuning a pre-trained model with down-sized data for the new scenario. Here, the main questions in performing cross-scenario PMP are: (1) How quickly and with how minimal data PMNet can effectively adapt to new scenarios; and (2) Which pre-trained model should be utilized for optimal performance in cross-scenario PMP.
}

\subsubsection{Efficiency}
For cross-scenario PMP, rapidly adapting PMNet models to new network scenarios using limited data is essential due to the time-consuming and expensive nature of channel measurement using RT simulation or channel sounding. This is particularly critical for applications like beam management and localization using ML-based PMP, which demand quick adjustments for new scenarios.

\begin{figure}[ht!]
\vspace{-0.5em}
\centering
\begin{subfigure}[t]{.48\linewidth}
\centering  
\includegraphics[width=\linewidth]{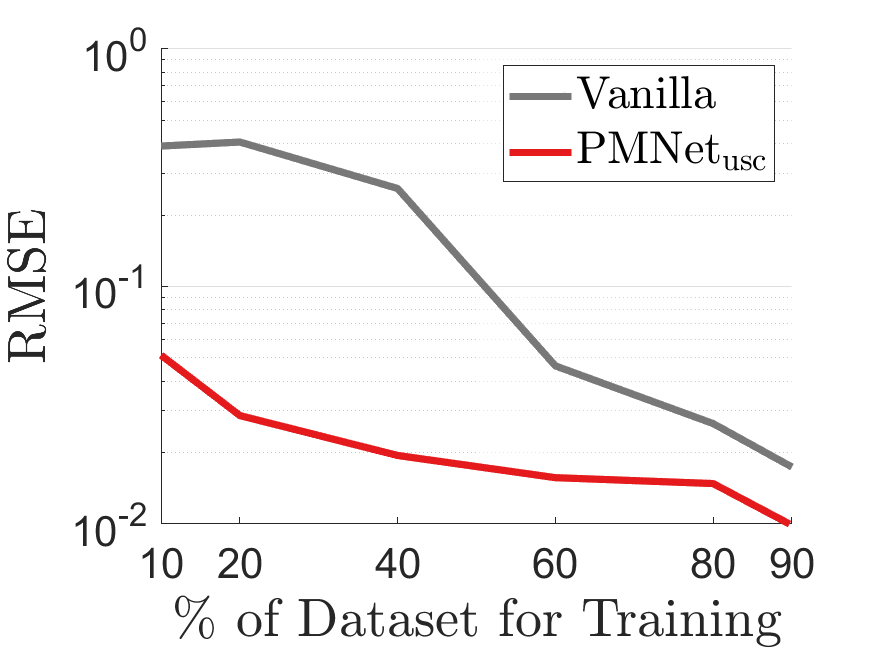}
\caption{RMSE}
\end{subfigure}\hspace*{.015\textwidth}%
\begin{subfigure}[t]{.48\linewidth}
\centering
\includegraphics[width=\linewidth]{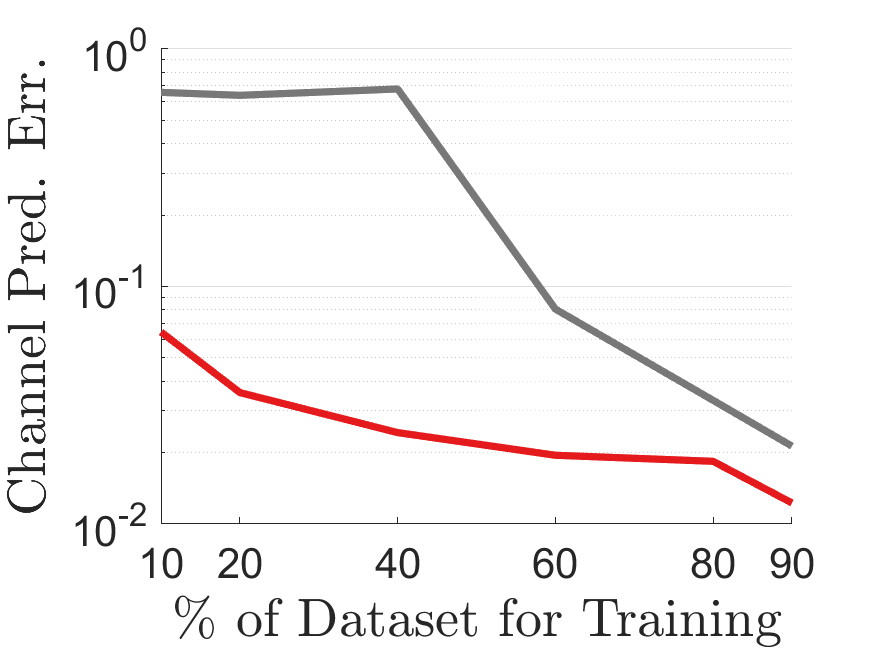}
\caption{Channel prediction error}
\end{subfigure}
\caption{Impact of TL on training data requirements. PMNet models with or without PMNet$_{\mathrm{usc}}$ pre-trained model are trained with 50 epochs and evaluated on the Boston dataset.}
\label{fig:comparison_TL_data}
\end{figure}

\tblue{
\BfPara{Impact of TL}\quad
TL can significantly improve the training speed of PMNet models for cross-scenario PMP.
As shown in Fig. \ref{fig:comparisn_TL_val} and Table \ref{table:requiredStep}, the TL case with the PMNet$_{\mathrm{usc}}$ pre-trained model achieves a given level of accuracy much faster even with much less amount of training data.
In particular, PMNet$_{\mathrm{usc}}$ achieves the same level of accuracy (\textit{RMSE}~$\leq 0.1$ and \textit{RMSE}~$\approx 0.03$) $\times 5.6$ and $\times 4.1$ faster, respectively, as the Vanilla case (highlighted in \circled{C}), where we define as ''Vanilla" the training from scratch in a particular environment. 

\begin{table}[ht!]
\centering          
\resizebox{1.\columnwidth}{!}{\begin{minipage}[h]{1.13\columnwidth}
\centering
\newcolumntype{R}{>{\raggedleft\arraybackslash}X}
\begin{tabularx}{1\linewidth}{l l l }
\toprule[1.5pt]
\multirow{2}{*}{\textbf{\textit{Case}}} & \multicolumn{2}{c}{\textbf{\textit{\# of Required Step (Training Speed)}}}\\    
\cmidrule(lr){2-3}
& \ \ \ \ \ \textbf{\textit{RMSE~$\leq 0.1 $}} & \ \ \ \ \ \textbf{\textit{RMSE~$\approx 0.03$}}\\     
\cmidrule(lr){1-1} \cmidrule(lr){2-2} \cmidrule(lr){3-3}
Vanilla ($90\%$ Data) 
& {5841} \hspace{-0.5pt}  
\tikz{\fill[fill=color7] (0.0,0) rectangle (1.3,0.2); \fill[pattern=north west lines, pattern color=black!30!color7] (0.0,0) rectangle (1.3,0.2);} ($\mathbf{\times 1.0}$) 
& {6195} \hspace{-0.5pt}  
\tikz{\fill[fill=color7] (0.0,0) rectangle (1.45,0.2); \fill[pattern=north west lines, pattern color=black!30!color7] (0.0,0) rectangle (1.45,0.2);} ($\mathbf{\times 1.0}$) \\
PMNet$_{\mathrm{usc}}$ ($20\%$ Data) 
& 1040 \hspace{-0.5pt} 
\tikz{\fill[fill=color1] (0.0,0) rectangle (0.3,0.2);\fill[pattern=north west lines, pattern color=black!30!color1] (0.0,0) rectangle (.3,0.2);} ($\mathbf{\times 5.6}$)
& 1520 \hspace{-0.5pt}  \tikz{\fill[fill=color1] (0.0,0) rectangle (0.45,0.2); \fill[pattern=north west lines, pattern color=black!30!color1] (0.0,0) rectangle (.45,0.2);} ($\mathbf{\times 4.1}$) \\
\bottomrule[1.5pt]
\end{tabularx}

\end{minipage}}
\caption{Impact of TL on training speed ($=\frac{1}{\mathrm{steps}}$). PMNet models with or without PMNet$_{\mathrm{usc}}$ pre-trained model are trained and evaluated on the Boston dataset.}
\label{table:requiredStep}
\end{table}

Furthermore, the TL can also significantly save the required amount of data for cross-scenario PMP. As shown in Fig. \ref{fig:comparison_TL_data}, the TL (PMNet$_{\mathrm{usc}}$) trained with about $20 \%$ of the Boston dataset achieves equivalent results to the Vanilla case trained with about $90 \%$ of the dataset. 

It is worth noting that limited training data can easily induce overfitting, as observed in the Vanilla case with $20 \%$ Data (highlighted in \circled{D}). For the same amount of new scenario data, the TL case (PMNet$_{\mathrm{usc}}$ ($20 \%$)) does not experience the overfitting issue. This suggests that TL also enhances training stability (less overfitting issue with limited data) in cross-scenario PMP.

Our findings demonstrate that the pre-trained PMNet$_{\mathrm{usc}}$ model efficiently accelerates the training process by leveraging its knowledge of PMP tasks, including the physics of wireless channel propagation and RoI segmentation, and this model can be readily adapted to new scenarios with minimal data and training steps.

Consequently, we confirm that fine-tuning with a \emph{suitable} pre-trained model is an effective cross-PMP task method. Another key question is which pre-trained model is suitable and which is not, which is discussed further in the following.
}

\begin{table*}[h!t]
\centering
\resizebox{.86\linewidth}{!}{\begin{minipage}{.78\linewidth}
\centering
\begin{tabularx}{1\linewidth}{l || l l ||c c c}
\toprule[1.5pt]
\textit{\textbf{Case}} & \ \  \textit{\textbf{Pre-training}} &  \ \textit{\textbf{Model}} & \textit{\textbf{RMSE}$\downarrow$} & \textit{\textbf{RoI Segmentation Err.}$\downarrow$} & \textit{\textbf{Channel Prediction Err.}$\downarrow$}  \\
\cmidrule(lr){1-1} \cmidrule(lr){2-2} \cmidrule(lr){3-3} \cmidrule(lr){4-4} \cmidrule(lr){5-5} \cmidrule(lr){6-6} 

Vanilla & \ \ \xmark & PMNet & 0.03415 & 0.02935 & 0.03844
\\
TL (ImageNet) & \ \  \cmark \ (ImageNet) & VGG16 & 0.04528 & 0.01814 & \ \ \ \ 0.05108 \circled{E}
\\
TL (3GPP) & \ \  \cmark \ (3GPP) & PMNet & 0.02809 & \textbf{0.00655} & 0.03238
\\
TL (USC) & \ \  \cmark \  (USC) & PMNet & \textbf{0.02792} & 0.01666 & \textbf{0.03145}
\\
\bottomrule[1.5pt]
\end{tabularx}





\vspace{-0.17em}
\subcaption{UCLA}
\vspace{0.7em}
\begin{tabularx}{1\linewidth}{l || l l ||c c c}
\toprule[1.5pt]
\textit{\textbf{Case}} & \ \  \textit{\textbf{Pre-training}} &  \ \textit{\textbf{Model}} & \textit{\textbf{RMSE}$\downarrow$} & \textit{\textbf{RoI Segmentation Err.}$\downarrow$} & \textit{\textbf{Channel Prediction Err.}$\downarrow$}  \\
\cmidrule(lr){1-1} \cmidrule(lr){2-2} \cmidrule(lr){3-3} \cmidrule(lr){4-4} \cmidrule(lr){5-5} \cmidrule(lr){6-6} 

Vanilla & \ \ \xmark & PMNet & 0.01736 & 0.02417 & 0.02125
\\
TL (ImageNet) & \ \  \cmark \ (ImageNet) & VGG16 & 0.01999 & \textbf{0.02040} & 0.02512
\\
TL (3GPP)  & \ \  \cmark \ (3GPP) & PMNet & 0.01762 & 0.04030 & 0.02187
\\
TL (USC)  & \ \  \cmark \  (USC) & PMNet & \textbf{0.00987} & 0.03530 & \textbf{0.01225}
\\
\bottomrule[1.5pt]
\end{tabularx}




\vspace{-0.17em}
\subcaption{Boston}
\end{minipage}}
\caption{Comparison of pre-trained models (VGG16$_{\mathrm{ImgNet}}$, PMNet$_{\mathrm{3gpp}}$, and PMNet$_{\mathrm{usc}}$) in terms of accuracy. Models are evaluated on the UCLA and Boston datasets, using $90 \%$ of the data for training and $10 \%$ of the data for validation. $50$ epochs are used for training. Lower values indicate better performance, and the lowest errors are highlighted.}
\label{table:comparison_TL}
\end{table*}

\begin{figure*}[!ht]
\centering
\begin{subfigure}[t]{.19\textwidth}
\centering
\includegraphics[width=\linewidth]{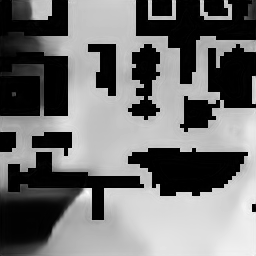}
\caption{Vanilla (UCLA)}
\end{subfigure}\hspace*{.015\textwidth}%
\begin{subfigure}[t]{.19\textwidth}
\centering
\includegraphics[width=\linewidth]{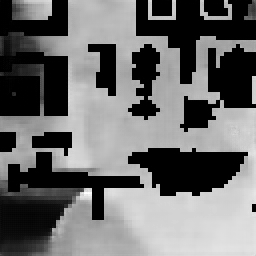}
\caption{VGG16$_{\mathrm{ImgNet}}$ (UCLA)}
\end{subfigure}\hspace*{.015\textwidth}%
\begin{subfigure}[t]{.19\textwidth}
\centering
\includegraphics[width=\linewidth]{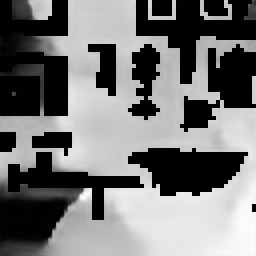}
\caption{PMNet$_{\mathrm{3gpp}}$ (UCLA)}
\end{subfigure}\hspace*{.015\textwidth}%
\begin{subfigure}[t]{.19\textwidth}
\centering
\includegraphics[width=\linewidth]{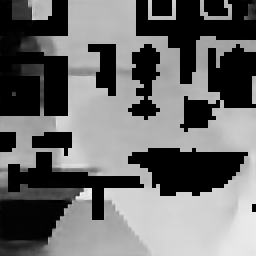}
\caption{PMNet$_{\mathrm{usc}}$ (UCLA)}
\end{subfigure}\hspace*{.015\textwidth}%
\begin{subfigure}[t]{.19\textwidth}
\centering
\includegraphics[width=\linewidth]{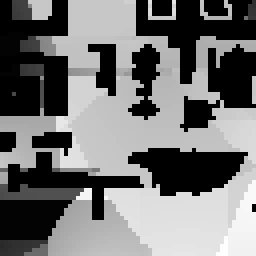}
\caption{Ground-truth (UCLA)}
\end{subfigure}
\\ \vspace{1.em}
\begin{subfigure}[t]{.19\textwidth}
\centering
\includegraphics[width=\linewidth]{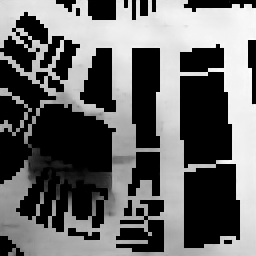}
\caption{Vanilla (Boston)}
\end{subfigure}\hspace*{.015\textwidth}
\begin{subfigure}[t]{.19\textwidth}
\centering
\includegraphics[width=\linewidth]{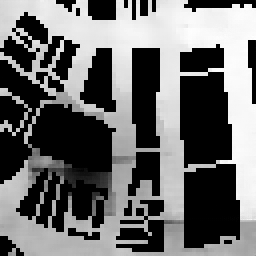}
\caption{VGG16$_{\mathrm{ImgNet}}$ (Boston)}
\end{subfigure}\hspace*{.015\textwidth}%
\begin{subfigure}[t]{.19\textwidth}
\centering
\includegraphics[width=\linewidth]{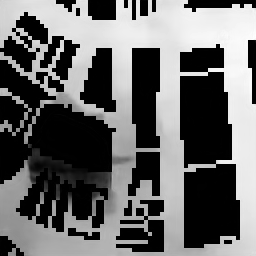}
\caption{PMNet$_{\mathrm{3gpp}}$ (Boston)}
\end{subfigure}\hspace*{.015\textwidth}%
\begin{subfigure}[t]{.19\textwidth}
\centering
\includegraphics[width=\linewidth]{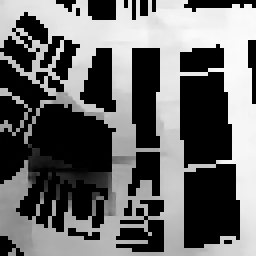}
\caption{PMNet$_{\mathrm{usc}}$ (Boston)}
\end{subfigure}\hspace*{.015\textwidth}%
\begin{subfigure}[t]{.19\textwidth}
\centering
\includegraphics[width=\linewidth]{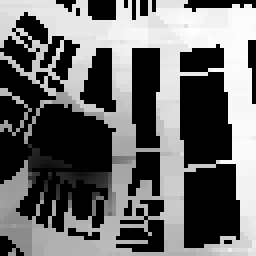}
\caption{Ground-truth (Boston)}
\end{subfigure}
\caption{Comparison of the prediction results of pre-trained models (VGG16$_{\mathrm{ImgNet}}$, PMNet$_{\mathrm{3gpp}}$, and PMNet$_{\mathrm{usc}}$). All models are trained using 50 epochs.}
\label{fig:comparison_TL}
\vspace{-1.em}
\end{figure*}

\subsubsection{Accuracy}
\tblue{
As discussed in Sec. \ref{sec:transferlearning}, the source and target scenario (task or domain) should be sufficiently similar for effective TL to occur. For instance, to successfully apply TL to the target task of predicting wireless communication channels, the NN should extract relevant features of wireless propagation physics from the source task. 

\BfPara{``Suitable" pre-trained model}\quad
Table \ref{table:comparison_TL} compares the performance of the PMNet model with and without TL. The baseline model, referred to as Vanilla, is trained without any TL (without any pre-trained model). Additionally, we compare the performance of TL using a pre-trained model trained on an unrelated source scenario (\textit{i.e.}, VGG16 trained on ImageNet) with TL using a pre-trained model trained on a related source scenario (\textit{i.e.}, PMNet trained on USC or 3GPP datasets).

As shown in Table \ref{table:comparison_TL}, both PMNet models trained on PMNet$_{\mathrm{usc}}$ and  PMNet$_{\mathrm{3gpp}}$ outperform the Vanilla case on all performance metrics, suggesting that using a pre-trained model trained on a related source task can significantly improve accuracy.

Interestingly, while the VGG16 model trained on ImageNet (VGG16$_{\mathrm{{ImgNet}}}$) outperforms the Vanilla for RoI segmentation, it fails to do so for channel prediction (highlighted in \circled{E}). This discrepancy stems from the VGG16 pre-trained model, which has an inherent understanding of segmentation and image representation from its source task; however, does not have any knowledge of the physics of wireless propagation. 

Fig. \ref{fig:comparison_TL} visually confirms the findings from Table \ref{table:comparison_TL}. All models achieve high accuracy for RoI segmentation, while only the TL case using a pre-trained model trained on a related source scenario (\textit{e.g.}, PMNet$_{\mathrm{3gpp}}$ and PMNet$_{\mathrm{usc}}$) achieves high accuracy for channel prediction, capturing subtle details of the wireless propagation physics. 
This suggests that our PMNet pre-trained model is generalizable to different scenarios with its inherent knowledge of channel propagation representation and that TL can further improve accuracy.

These results empirically demonstrate that pre-trained model's source dataset (task or domain) should be similar to the target dataset (task or domain) to transfer useful information during TL. Specifically, for cross-scenario PMP, it is important to use a pre-trained model that has been trained extensively on data related to wireless propagation physics.

Therefore, we conclude that the suggested TL approach, fine-tuning with a stable and closely related pre-trained model (such as PMNet$_{\mathrm{usc}}$), is a simple yet effective way to address the cross-scenario PMP task, which is important for practical applications. 
}





\section{Conclusion}
\label{sec:conclusion}
This work introduces an ML-based large-scale channel prediction framework, PMNet, which can create highly accurate pathloss predictions for a given map in a few milliseconds. 
Utilizing an RT channel measurement dataset of real-world scenarios (\textit{e.g.}, USC, UCLA, and Boston area), PMNet is verified for its accuracy and training efficiency.
In particular, TL with our PMNet pre-trained model, which has generalization capability for different network scenarios, enables the PMNet to adapt itself quickly and efficiently to a new network scenario, while achieving an RMSE of $10^{-2}$ level. 

The high accuracy and low runtime of the PMNet framework make it suitable for deployment planning in dense networks as well as online optimization of network parameters. 

\tblue{
Still, it remains an open question whether the knowledge of wireless propagation physics in our PMNet pre-trained model can be transferred to other downstream tasks beyond the PMP task; this question will be the topic of our future research.
}


\bibliographystyle{IEEEtran} 
\bibliography{refs}

\end{document}